\documentclass[11pt, final]{myarticle} 
\usepackage{setspace}
\usepackage{amssymb,cite,latexsym,amsmath,subfigure}
\usepackage{graphicx}
\usepackage{psfrag}
\usepackage{pstricks}
\makeatletter
\renewcommand\@seccntformat[1]{\csname the#1\endcsname.\quad}

\date{}
\begin{document} 

\title{{\Large Green's function-stochastic methods framework for
probing nonlinear evolution problems: Burger's equation, the nonlinear
Schr\"odinger's equation, and hydrodynamic 
organization of near-molecular-scale vorticity}} 
\author{R. G. Keanini\footnote{
Address: Mechanical Engineering \& Engineering Science, UNC-Charlotte,
9201 University City Blvd, Charlotte, NC 28223-0001.
Email: rkeanini@uncc.edu; Phone: 704-687-8336; Fax: 704-687-8345}\\
Department of Mechanical Engineering and Engineering Science\\
The University of North Carolina at Charlotte \\
Charlotte, North Carolina 28223-0001}
\maketitle
\begin{abstract}
A framework which combines Green's function (GF) methods and
techniques from the theory of stochastic processes 
is proposed for tackling 
nonlinear evolution problems.
The framework, established by a series of easy-to-derive
equivalences between Green's function and stochastic representative 
solutions of
\textit{linear} drift-diffusion problems,
provides a flexible structure within which
nonlinear
evolution problems
can be analyzed and physically probed.
As a preliminary test bed, 
two canonical, nonlinear evolution problems -
Burgers' equation and the nonlinear Schr\"odinger's
equation - are first treated.
In the first case, the framework provides a rigorous,
probabilistic derivation of the well known
Cole-Hopf ansatz. Likewise, in the second, the machinery
allows systematic recovery of a known soliton solution.
The framework
is then applied to a fairly extensive
exploration of physical features underlying
evolution of randomly stretched and advected 
Burger's vortex sheets.
Here, the governing vorticity equation corresponds to the Fokker-Planck equation
of an Ornstein-Uhlenbeck process, a
correspondence that motivates an investigation of sub-sheet
vorticity evolution and organization.
Under the assumption that weak hydrodynamic fluctuations
organize disordered, near-molecular-scale, sub-sheet
vorticity, 
it is shown that
these modes consist of two weakly damped counter-propagating
cross-sheet acoustic modes, a diffusive cross-sheet
shear mode, and a diffusive cross-sheet entropy mode.
Once a consistent picture of in-sheet vorticity
evolution is established, 
a number of analytical results, 
describing the motion and spread of single, multiple, 
and continuous sets of Burger's vortex sheets, evolving
within deterministic and random strain rate fields,
under both viscous and inviscid conditions, are obtained.
In order to promote application to other nonlinear problems,
a tutorial development of the framework
is presented.
Likewise, time-incremental solution approaches and
construction of approximate, though otherwise
difficult-to-obtain backward-time GF's (useful 
in solution
of forward-time evolution problems) are discussed.
\end{abstract}

\vspace{.4cm}


\section{Introduction}
Parabolic evolution equations of the form
\begin{equation}\label{nonlinearpattern}
A (\mathbf{x}, \tau, \boldsymbol{\eta}) \boldsymbol{\eta}_{\tau} + B (\mathbf{x}, \tau,
\boldsymbol{\eta} ) 
\cdot \boldsymbol{\nabla} \boldsymbol{\eta} + C(\mathbf{x}, \tau, \boldsymbol{\eta}) 
\nabla^2 \boldsymbol{\eta} =
\mathbf{F}(\mathbf{x}, \tau, \boldsymbol{\eta})
\end{equation}
lie at the heart of a far-reaching set of physical theories and models.
Here, $ \boldsymbol{\eta} , $ the
variable of interest,
and $ A , $ $ B , $ $ C , $ and $ F $ can 
represent scalar, vector, or higher order tensor
quantities, and 
$ \mathbf{x} $ and $ \tau $ correspond respectively to 
space- and time-like variables.
A limited list
of examples include
the Navier-Stokes equations (see, e.g., \cite{batchelor}), 
Schr\"odinger's equation
\cite{messiah}, advection-diffusion
equations \cite{deen}, reaction-diffusion equations \cite{williams}, equations
governing nonlinear pattern formation \cite{grindrod}, a diverse set 
of wave equations
\cite{whitham, drazin}, 
Chapman-Kolmogorov equations
governing evolution of Markov processes \cite{gardiner},
and particle and continuum versions of mass, charge, and
linear and angular momentum conservation \cite{forsterbook, hughes}.

Finding general approaches for treating \textit{nonlinear} 
versions of (\ref{nonlinearpattern})
drives a large field of research \cite{ablowitz, ablowitz2,
wang, walquist, abdou, camassa, honerkamp}. 
Numerical methods are typically 
favored in applied problems, while
intense interest attaches to development of analytical techniques
since these can 
deepen physical and mathematical insight \cite{ablowitz,walquist,
camassa, honerkamp}, and somewhat secondarily,
allow, e.g., code benchmarking and 
algorithm development \cite{taigbenu, drazin}. 

Green's function (GF) methods, of course, find wide
application to \textit{linear} \eqref{nonlinearpattern},
where the method's versatility derives from several features: i) 
GF solution structure
exemplifies simplicity,
providing a transparent description of a linear system's 
response to often complicated space- and history-dependent
boundary and initial conditions, and internal forcing;
ii) GF's employ, via delta functions, space- and time-localized, but
physically-lumped 
i.e., \textit{black-box}, descriptions of typically
ill-understood system interactions with these forcing agents;
and iii) GF's can often be ascribed intuitive interpretations
\cite{economou,feynmankatzref, ashcroft, jackson, deen, kimura},
promoting physical understanding.

A number of papers have used Green's function approaches
to tackle \textit{nonlinear} evolution problems \cite{ablowitz,taigbenu,arkin}. 
However, 
with the exception of a limited collection
of \textit{systematic} techniques, e.g.,
Green's element methods \cite{taigbenu} and
the inverse scattering transform \cite{ablowitz}, 
work in this area remains in an early state of
development.

A well-known, though little-used \textit{stochastic}
representative solution for the non-homogeneous
Cauchy problem, a linear drift-diffusion embodiment
of \eqref{nonlinearpattern},
was presented in 1931 by Kolmogorov \cite{kolmogorov, friedman}. 
In analogy with the Green function's circumvention
of unresolved small-scale forcing,
the stochastic solution
rests on Wiener's \cite{gardiner} physically lumped
description of unresolved \textit{random} forcing.
The latter feature, paralleling the role played by
delta functions in GF models,
promotes tremendous versatility in 
both the reach and interpretation of
stochastic process models.

This paper pursues four objectives:\\
\noindent \textbf{A.} Given the \textit{linear} nonhomogeneous form 
of equation 
(\ref{nonlinearpattern}),
subject to nonhomogeneous initial and boundary conditions,
Kolmogorov's stochastic representative solution can be expressed
\cite{kolmogorov, friedman}.
Our first objective 
centers on highlighting a series of equalities
that exist between expectations appearing in Kolmogorov's
solution and corresponding terms in the
Green's function solution. This demonstration is significant
since it establishes a simple
\textit{bridge} between two broad fields, Green's function methods and the
theory of stochastic processes. 

Importantly, recognition
of this connection
establishes a 
structure in which stochastic and Green's function 
methods can be applied in concert to a range of
deterministic and random evolution problems
of the form in \eqref{nonlinearpattern}.
We refer to this structure as 
the \textit{Green's function-stochastic methods framework} (GFSM).

\noindent \textbf{B.} Given this framework, the remainder of the
paper focuses on two principal questions: 
\renewcommand{\labelenumi}{\alph{enumi})}
\begin{enumerate}
\item Can the framework be \textit{systematically} applied
to solve \textit{nonlinear} versions of 
equation \eqref{nonlinearpattern}?
\item Can the framework facilitate physical and mathematical
exploration of 
problems characterized by some element of randomness?
\end{enumerate}

With regard to the first question,
since most nonlinear problems require some form of
numerical attack, our second objective 
focuses on set-up of time-incremental solutions.
The main ideas include: i) derivation of exact, space-dependent
Green's functions, valid over arbitrary, though small time
increments, ii) identification of forward and backward time
evolution problems and associated adjoint problems, and iii)
as illustrated in Test Case 1, use of approximate
forward-time Green's functions as surrogates for hard-to-compute
backward time GF's.

\noindent \textbf{C.} The third objective, also addressing question a), 
centers on \textit{testing}
the viability of time-incremental attacks
on nonlinear versions of \eqref{nonlinearpattern}.
Two nonlinear evolution equations having
exact solutions,
Burger's equation and the nonlinear Schr\"odinger equation, provide test beds.
In both Test Cases, we start from a general,
time-incremental Green's function solution, and identify strategies
for obtaining known non-incremental solutions.

\noindent \textbf{D.} The last objective, addressing question b),
aims at illustrating how GF and stochastic process ideas can provide essential
physical guidance in the analysis
of nonlinear evolution problems.
For this demonstration, and as presented in section 6,
we study 
a canonical fluid mechanics
problem which captures
the ubiquitous, combined effects of 
vortex stretching, advection, and diffusion:
evolution of Burger's vortex sheets. 

As described in section 6, much of
the analysis
pivots on gaining an understanding 
of near-molecular scale, sub-sheet
vorticity transport and organization. 
Crucially, this question emerges when we observe 
that evolution of sheet-scale vorticity, initiated by 
a delta function initial condition, corresponds to 
evolution of the transition density for an Ornstein-Uhlenbeck
process. 

In other words, in making this connection,
we are presented with a radically alternative, probabilistic picture:
development of any given Burger's vortex sheet, evolving within
either a deterministic or random strain rate
field, can be \textit{interpreted} as the collective,
stochastic evolution of a
swarm of \textit{elemental vortex sheets} (EVS).

Thus, in order to identify 
a reasonable physical embodiment of EVS's,
we are lead to investigate:
i) the
highly disorganized structure
of sub-sheet, short-time-scale vorticity,
and ii) the long BVS-time-scale organization of this vorticity.

Mathematically, couching analysis of BVS evolution
in terms of the stochastic evolution of elemental vortex sheets
likewise proves 
advantageous since it allows straightforward,
probabilistic determination of time-dependent
BVS mean position and spread.
Similarly, in the case of continuous initial vorticity distributions,
a stochastic vantage point leads naturally to
Feynman-Kac solutions for the resulting vorticity evolution.

In closing the Introduction, and as an aid to navigating the paper,
we highlight the paper's essential five-part structure:
\renewcommand{\labelenumi}{\Roman{enumi})}
\begin{enumerate}
\item Mathematical details needed for application of the
framework to linear and nonlinear evolution
problems are given in sections 2 and 3.
\item Testing use of \textit{time-incremental} GF's for
solving nonlinear problems is described in sections
4 and 5. 
\item Application
of Green's function and stochastic
process ideas as physical and mathematical probes is, as mentioned,
illustrated in section 6. 
The paper's final three parts
correspond to three distinct elements comprising the
illustration.
Thus, set-up and calculation of single sheet
GF's 
is carried out in sections 6.1 and 6.2.
\item Sections 6.3 shows that evolution
of \textit{individual} Burger's sheets
can be interpreted as the stochastic
evolution of a \textit{swarm} of sub-sheet elemental vortex sheets,
the latter representing a quasi-physical embodiment
of an underlying Ornstein-Uhlenbeck process.  Section 6.4
pursues a physical interpretation of this observation,
and in the process, addresses the fundamental
question of how highly disorganized, near-molecular-scale vorticity
becomes organized on long BVS length and time scales.
\item Finally, sections 6.5 through 6.7 investigate
the evolution of single, multiple, and continuous
collections of Burger's vortex sheets advected and stretched by
random
strain rate fields.
\end{enumerate}
%%%%%%%%%%%%%%%%%%%%%%%%%%%%%%%%%%%%%%%%%%%%%%%%%%%%%%%%%
\section{Green's function-stochastic methods framework}
As mentioned, the GFSM framework can be viewed as a union of 
Green's function methods
and the theory of stochastic processes,
\textit{bridged} by the equalities
\eqref{eqty1} through \eqref{eqty3} derived below.
This section focuses on 
derivation of 
equations \eqref{eqty1} through \eqref{eqty3}.  Section \ref{sec3}
and the Appendix describe time-incremental application of the framework to
nonlinear problems.

Although it is likely that the 
approach used here -- 
derivation of the Green's function
solution  
followed by comparison with the backward-time
stochastic solution --
has been described elsewhere,
we have not located such descriptions. We note that
Friedman \cite{friedman}, following Kolmogorov \cite{kolmogorov},
derived an essential result:
the representative stochastic solution
of a backward time, linear, nonhomogeneous advection-diffusion problem.
However, \cite{kolmogorov, friedman} did not
connect the stochastic solution to
the equivalent Green's function 
solution.

\noindent Here, we use
the following non-rigorous recipe:
\renewcommand{\labelenumi}{\roman{enumi})}
\begin{enumerate}
\item Define an appropriate forward time 
linear evolution problem. 
\item Define the associated adjoint
problem governing the Green's function, $ G . $  
\item Use the adjoint problem to derive the 
Green's function solution. 
\item Write the linear evolution problem in backward time form
and recognize that the result is 
a backward Kolmogorov (Fokker-Planck) equation;
thus, express the solution as a representative stochastic solution
\cite{friedman}. 
\item Finally, equate corresponding terms in both solutions,
i.e., equate terms involving 
boundary conditions, the initial condition, and the
nonhomogeneous forcing term.
\end{enumerate}
\noindent Remark 1: It is important to note 
that for divergence-free drift fields,
the validity of this approach can be easily
proven: under these circumstances,
the adjoint problem governing  $ G ,$ 
and the backward time Fokker-Planck problem
governing the transition density, $ p , $ are identical. Thus, the 
Green's function
can be interpreted as a transition density, 
and hence the Green's function-stochastic solution equalities 
in \eqref{eqty1} - \eqref{eqty3} 
become 
identities. 

\noindent Remark 2: For application of the 
GFSM to \textit{nonlinear} evolution problems,
the \textit{appropriate linear} evolution equation
simply corresponds to a linearized version
of the original. As illustrated
in Test Cases 1 and 2, 
linearization takes advantage of the fact that
the solution is constructed time-incrementally.
%%%%%%%%%%%%%%%%%%%%%%%%%%%%%%%%%%%%%%%%%%%%%%%%%%%%%%%%
\subsection{Evolution problem template}
The examples treated in this paper 
can be mapped in some fashion,
e.g., time-incrementally for nonlinear problems, to the following
linear drift-diffusion problem: 
\begin{equation}\label{proc1}
M \eta = -f (\mathbf{x'} , t') \hspace{1.2cm} 
\mathrm{on} \ \ Q = D \times (0,t] 
\end{equation}
\begin{equation}\label{proc1a}
\eta ( \mathbf{x'},0) = \phi(\mathbf{x'})  \hspace{1.2cm} \mathrm{on} \ \ 
\delta Q_o = D \times \{t' = 0 \}
\end{equation}
\begin{equation}\label{proc1b}
\eta (\mathbf{x'},t' ) = g(\mathbf{x'}, t' )  \hspace{1.2cm} 
\mathrm{on} \ \ \ \delta Q = \delta D \times (0,t]
\end{equation}
where the operator, $ M , $ 
is given by
\begin{equation}\label{moper}
M \eta = \nu \nabla^{'2} \eta  + 
\mathbf{b} \cdot \boldsymbol{\nabla'} \eta  - \frac{\partial \eta }{\partial t' }
\end{equation}
and where, for notational convenience, we
use the backward drift $ \mathbf{b} $ in place of the forward
(actual) drift $ \mathbf{v} $ (with
$ \mathbf{b} = - \mathbf{v} ) .$ In addition, $ \phi (\mathbf{x'}) $ is the initial
condition,
$ g (\mathbf{x'} , t' ) $ is a time-varying Dirichlet condition,
the differential
operators $ \nabla^{'2} $ and $ \boldsymbol{\nabla'} $
denote derivatives taken with respect to $ \mathbf{x'} , $
and $ D $ is the spatial solution domain. 
See the Appendix for a description of forward and backward time coordinates
and forward and backward evolution equations.
%%%%%%%%%%%%%%%%%%%%%%%%%%%%%%%%%%%%%%%%%%%%%%%%%%%%%%%%%%%%%%5
\subsection{Representative stochastic solution}
The \textit{backward} time stochastic solution of the 
evolution problem in (\ref{proc1})-(\ref{moper})
within a 
space-time domain, $ Q = D \times [s , T ) , $ subject to Dirichlet
conditions, $ \eta = g(\mathbf{x}, s' ) , $ on the boundary $ \delta Q $ of $ Q , $
and a final condition $ \eta = \phi (\mathbf{x} , s = T )  $
on the final time-slice, $ D \times \{ s = T \} ,$
can be expressed in representative form as
\cite{friedman, kolmogorov}:
\begin{equation}\label{sol1}
\eta (\mathbf{x}, s) = 
E_{ \mathbf{x}, s} \big[ g \big( \boldsymbol{\chi}(\tau) , \tau \big) \big]
+ E_{ \mathbf{x}, s} \big[ \phi \big( \boldsymbol{\chi} (T) \big) \big]
+ E_{ \mathbf{x}, s} \big[ \int_{s}^{\tau}
f \big( \boldsymbol{\chi} (s'), s' \big) ds' \big]
\end{equation}
where $ s = T - t , $ is 
backward time, 
$ t $ forward time, and
$ T $ is the backward time instant at which the field 
$ \eta  (\mathbf{x} , s ) $
is known. We outline a rough proof of this below.
Here, $ E_{\mathbf{x}, s} $ is the expectation associated with 
the stochastic process, $ \boldsymbol{\chi}(s) , $
sampling respectively, Dirichlet conditions on $ \delta Q , $ the final time condition,
$ \phi (\mathbf{x}, T) $ on $ D \times \{s=T \} , $ and the forcing function $ f $ within $  Q . $
Thus, $ \tau $ represents the random time at which the process meets (and is absorbed on)
the Dirichlet boundary.
%%%%%%%%%%%%%%%%%%%%%%%%%%%%%%%%%%%%%%%%%%%%%%%%%%%%%%%%
\subsection{Bridge relations between Green's function and stochastic solutions}
In order to clearly highlight the origin of the bridge relations,
we first outline well-known steps for generating the GF solution
to \eqref{proc1} - \eqref{moper}, and then present a non-rigorous,
though easy-to-grasp derivation of Kolmogorov's solution, equation
\eqref{sol1}.
%%%%%%%%%%%%%%%%%%%%%%%%%%%%%%%%%%%%%%%%%%%%%%%%
\subsection*{Green's function solution} 
The Green's function solution slightly extends the 
well-known derivation
outlined, e.g., by 
Morse and Feshbach \cite{morsefeshbach}
and Barton \cite{barton}; the extension consists
of incorporating non-zero drift,
$ \mathbf{v} = - \mathbf{b} $ into the governing equation for $ \eta . $ 

\renewcommand{\labelenumi}{\roman{enumi})}
\noindent The derivation, stated in tutorial fashion, is as follows:
\begin{enumerate}
\item Given the forward time operator $ M $ defined in (\ref{moper}),
find the corresponding adjoint operator $ M^* $
(see, e.g., \cite{arfken}),
here given by \cite{friedman}:
\begin{equation}\label{proc1adj}
M^* = \nu {\nabla'}^2  - \mathbf{b} \cdot \boldsymbol{\nabla'} - 
\boldsymbol{\nabla'} \cdot \mathbf{b}  
+ \frac{\partial }{\partial t' }
\end{equation}

\item Define a Green's function, $ G (\mathbf{x} , t | \mathbf{y} , t')  , $ 
%s\nu \int_0^t \oint_{S} \eta \boldsymbol{\nabla'} \Gamma \cdot
%\mathbf{n'} dS' dt'
satisfying 
\begin{equation}\label{gammaeqn}
M^* G = - \delta (t - t') \delta (\mathbf{x} - \mathbf{y})
\end{equation}
for all 
$ \mathbf{x}, \mathbf{y} \in D , $ and $ 0 \leq t' \leq t  . $  

\item Multiply (\ref{proc1}) by $ G (\mathbf{x} , t | \mathbf{x'} , t') $ and 
(\ref{gammaeqn}) by $ \eta (\mathbf{x'} , t') , $ 
and form:
\begin{equation}\label{greenthm}
\int_{0}^{t+\epsilon} \int_D [ G ( M \eta ) - \eta (M^* G) ] dD' dt' =
- \int_{0}^{t+\epsilon} \int_D [ G f - \eta \delta(\mathbf{x} - \mathbf{x'}) 
\delta ( t - t') ] dD' dt'
\end{equation}
which allows application of Green's theorem.
Here, $ \epsilon / t << 1 , $ and $ G = 0 $ for $ t' >  t . $ 

\item Carry out the integrations
in (\ref{greenthm}) to obtain the solution
for $ \eta (\mathbf{x} , t) : $
\begin{equation}\label{greensoln}
\eta(\mathbf{x} , t ) = \int_0^t \int_D G f dD'  dt' -
\nu \int_0^t \oint_{\delta D} \eta \boldsymbol{\nabla'} G \cdot
\mathbf{n'} dS' dt' +
\int_{D} \phi (\mathbf{x'}) G (\mathbf{x}, t | \mathbf{x'} , 0) dD' -
\int_0^t \oint_{\delta D} G \mathbf{J} \cdot \mathbf{n'} dS' dt'
\end{equation}
where $ \mathbf{J} = \mathbf{v} \eta -
\nu \boldsymbol{\nabla'} \eta  $ is the total flux of $ \eta $ (and where again $ \mathbf{b} = - 
\mathbf{v} ) . $ 
\end{enumerate}
%%%%%%%%%%%%%%%%%%%%%%%%%%%%%%%%%%%%
\subsection*{Stochastic solution}
This derivation, again presented in recipe fashion,
is a \textit{wholly heuristic,} non-rigorous version of
that given by, e.g., Friedman \cite{friedman}. 
A non-technical approach is preferred since it provides 
a simple path to the desired result, equation \eqref{sol1}.
\renewcommand{\labelenumi}{\roman{enumi})} 
\begin{enumerate}
\item Consider the differential, random, multidimensional
displacement 
of the stochastic process, 
$ \boldsymbol{\chi} , $ 
over the backward time interval $ ds' : $
\begin{equation}\label{chieqn}
d \boldsymbol{\chi}(s') = \mathbf{b} (\boldsymbol{\chi} (s'), s') d s' 
+ \sqrt{2 \nu} d \mathbf{w} (s')
\end{equation}
where $ \mathbf{w} $ is a multi-dimensional
Wiener process.
For any given realization of the random displacement,
$ d \boldsymbol{\chi}(s') , $  use a Taylor expansion about
the solution point, $ (\mathbf{x}, s ), $
to compute the associated change
in $ \eta : $
\begin{equation}\label{difff}
d \eta = \left[ \eta_{s'} ds' + \mathbf{b} \cdot \boldsymbol{\nabla'} \eta 
+  \nu {\nabla'}^2 \eta \right] ds' 
\end{equation}
where, in anticipation
of taking the expectation over the Wiener process, 
terms in $ d w_i = w_i (s + ds') - w_i (s) , $ are set to $ 0 $ 
and terms in $ d w_i d w_j $ are expressed as $ d s' \delta_{ij} . $
In addition, we keep $ ds' $ small enough that quadratic and higher order 
terms in $ ds' $ can be neglected. Finally, since $ dw_i $ is a 
zero-mean, gaussian random variable, expectations of terms 
involving $ d w_i^{2+n} $
are zero for $ n = 1,2, 3, ... $  See, e.g., Gardiner \cite{gardiner}
for further details.
\item Next, integrate \eqref{difff} along the 
random path traced by the process $ \boldsymbol{\chi}(s') $
as it progresses from $ s $ toward the backward time, $ T : $
\begin{equation}\label{integralf}
\eta(\mathbf{x},s) = \eta(\boldsymbol{\chi}(\tau), \tau) -
\int_s^{\tau} \left[ \eta_{s'} ds' + 
\mathbf{b} \cdot \boldsymbol{\nabla'} \eta 
+  \nu {\nabla'}^2 \eta \right] ds' 
\end{equation}
where $ T $ corresponds
to the forward-time initial instant (say, $ t=0 ),$
and where the desired term, $ \eta( \mathbf{x},s) , $
has been isolated. Again, in anticipation of taking
expectations, terms involving
$ dw_i , $ $ d w_i d w_j $ with $ i \ne j , $ and $ dw_i^{2+n} $
have been excluded.
Here, the time $ \tau $
depends on where $ \boldsymbol{\chi}(s') $ ends up: for those realizations
that impact the hyper-surface, $ \delta Q = \delta D \times [0,T) , $
enclosing the space-time solution domain, $ Q = D \times [0,T) , $
prior to reaching the final space-time slice, $ D \times \{ s'=T \} , $
$ \tau $ is the (random) time of impact. For those realizations that
survive without impacting $ \delta Q , $
$ \tau = T . $
\item Next, replace the argument in the last integral
with the right side of \eqref{proc1}, take expectations 
with respect to the process $ \boldsymbol{\chi} , $
and use the fact that 
$ E_{\mathbf{x},s} \eta(\boldsymbol{\chi} (\tau) , \tau) = E_{\mathbf{x},s}
g \left(\boldsymbol{\chi}(\tau), \tau \right) + 
E_{\mathbf{x},s} \phi \left(\boldsymbol{\chi}(T) \right) , $ to 
obtain the final result, equation 
\eqref{sol1}. 
\end{enumerate}

In order to obtain the bridge relations, we
simply recognize
that in linear problems,
$ \eta $ at any point $ (\mathbf{x}, s ) , $ as
determined by the representative stochastic solution
in (\ref{sol1}) \textit{must} correspond to $ \eta(\mathbf{x}, t ) , $
as computed via the Green's function solution in \eqref{greensoln}.
Thus, comparing similar terms in (\ref{sol1}) and (\ref{greensoln}), noting the
correspondence between the backward solution time, $ s , $ and the forward
solution time, $ t , $ we obtain the following:
\begin{equation}\label{eqty1}
E_{\mathbf{x}, s} \ g \big( \boldsymbol{\chi} (\tau ) \big) 
= - \nu \int_0^t \oint_{\delta D} g (\mathbf{x'}, t') \boldsymbol{\nabla '} 
G(\mathbf{x}, t| \mathbf{x'} , t') \cdot \mathbf{n} dS' dt'
\end{equation}
\begin{equation}\label{eqty2}
E_{\mathbf{x}, s} \ \phi( \boldsymbol{\chi} (T) ) = 
\int_D \phi (\mathbf{x'}) G (\mathbf{x} , t | \mathbf{x'} , 0 ) dD'
\end{equation}
\begin{equation}\label{eqty3}
E_{\mathbf{x}, s} \int_s^{\tau} f \big( \boldsymbol{\chi} (s'), s' \big) ds' 
= \int_0^t \int_D f (\mathbf{x'}, t') G(\mathbf{x}, t| \mathbf{x'} , t') d \mathbf{x'} dt'
\end{equation}
where (\ref{proc1b}) has been used in the second term on the right in (\ref{greensoln}).

Note that an equivalent form of the
the equality in (\ref{eqty2}), valid for Cauchy problems in
unbounded domains, is rigorously proven in Friedman \cite{friedman}.
Likewise, a relation equivalent to (\ref{eqty1}), and appropriate
to the Dirichlet problem $ M \eta = 0 , $ is given by Schuss \cite{schuss}.
In addition, in the case where drift $ \mathbf{b} = - \mathbf{v} $ is everywhere
zero, the solution in (\ref{greensoln}) is identical to that
given by Barton \cite{barton}.
Finally, stochastic representations are available for Neumann and mixed initial boundary
value problems; see, e.g., \cite{morillon}.
%%%%%%%%%%%%%%%%%%%%%%%%%%%%%%%%%%%%%%%%%%%%%%%
%%%%%%%%%%%%%%%%%%%%%%%%%%%%%%%%%%%%%%%%%%%%%%%
\section{Incremental solutions for nonlinear problems}\label{sec3} 
When confronting nonlinear and/or nonhomogeneous problems,
use of a time-incremental attack is immediately suggested. 
The idea is simple - shrink the forward solution time interval, $ \Delta t' $
(or in backward evolution problems, the backward interval $ \Delta s' ) $ 
in order to: i) allow linearization of nonlinear evolution problems, and ii)
when necessary, allow nonhomogeneous sources, 
$ \mathbf{F}(\mathbf{u_{j+1}}, \mathbf{x}, \tau_{j+1}) , $
to be expressed as
$ \mathbf{F}(\mathbf{u_{j}}, \mathbf{x}, \tau_{j}) , $
where in either forward or backward time, $ \tau_{j+1} = \tau_j + \Delta \tau. $

In this section, and 
in anticipation of 
Test Cases 1 and 2, we write down 
incremental versions of the stochastic 
and Green's function solutions
in (\ref{sol1}) and (\ref{greensoln}).
The Appendix describes essential requirements that
incremental GF's must meet.
%%%%%%%%%%%%%%%%%%%%%%%%%%%%%%%%%%%%%%%%%%%%%%%%%
\subsection{Incremental Green's function and stochastic solutions (1-D case)}
Test Cases 1 and 2 are initiated
from an incremental version of (\ref{greensoln}).
In both, we assume that
the solution point
$ (\mathbf{x} , t) \in  Q $ is sufficiently removed from 
boundaries to allow neglect of boundary conditions.
Thus, referring to (\ref{greensoln}), setting $ D = (-\infty , \infty) , $
and dropping boundary terms, we arrive at 
\begin{equation}\label{diffgreensoln}
\eta ( x , t_{j+1} ) = \int_{- \infty}^{\infty} 
G (x , t_{j+1} | x' , t_j ) \eta (x', t_{j} ) dx'
+ \int_{t_j}^{t_{j+1}} \int_{-\infty}^{\infty} G(x, t| x' , t') 
f (x', t') d x' dt'
\end{equation}
where, to allow incorporation of linearized nonlinear terms
as well as nonhomogeneous terms,
the nonhomogeneous source, $ f , $ is included.
Physically, \eqref{diffgreensoln} gives the response,
$ \eta(x, t_{j+1}) , $ at $ t_{j+1} $ due to both the 'initial condition',
$ \eta(x', t_j ) $ at $ t_j , $ and the time-dependent forcing, $ f(x', t') , $
that takes place over $ \Delta t_j = t_{j+1} -t_j. $

For completeness, we also write the associated 
time-incremental version of the representative stochastic solution in
(\ref{sol1}): 
\begin{equation}\label{diffsol1a}
\eta (x, t_{j+1} ) = E_{x, t_{j+1}} \eta ( \chi (t_{j}) , t_j ) - 
E_{x, t_{j+1}} \int_{t_{j+1}}^{t_j} f \big( \chi (s'), s' \big) ds' 
\end{equation}
where, for notational consistency, we have stated the backward time coordinate, say $ s_k , $ in terms of its
corresponding forward time coordinate, $ t_{j+1} $ (thus, for example, $ s_k \Leftrightarrow t_{j+1} $
and $ s_{k+1} \Leftrightarrow t_j ) . $ 

Finally, note that incremental versions of the
bridge relations in (\ref{eqty1})-(\ref{eqty3}) can be easily
written down by comparing like terms in, e.g., (\ref{diffgreensoln}) and (\ref{diffsol1a}).
%%%%%%%%%%%%%%%%%%%%%%%%%%%%%%%%%%%%%%%%%%%%%%%%%%%%%%%%%%%%%%
\section{Test Case 1: Solution of Burger's equation}\label{colehopfsec1}  
In brief overview, this example tests application of 
the GFSM framework to 
nonlinear drift-diffusion problems. As an aid to
understanding the development,
we note the following essential points:
\renewcommand{\labelenumi}{\roman{enumi})} 
\begin{enumerate}
\item A minimal requirement for 
application of time-incremental GF's to nonlinear problems
rests on derivation of an incremental GF.
By properly limiting the time step size,
the difficult-to-solve backward-form adjoint equation
can be recast in approximate, \textit{soluble,} forward form.
Section 4.2 assumes a simple stochastic process viewpoint
in order to derive generic time step constraints,
equations \eqref{timestepcond1} and \eqref{timestepcond2},
that must be met when using this procedure.
\item Nonlinear problems having a space or time-dependent 
drift, $ u , $ lead, not unexpectedly, to incremental
GF's containing the same. See \eqref{incgreensoln1} below.
As will be shown, it is sometimes possible to transform the
nonlinear GF to linear form by eliminating the \textit{a priori}
unknown $ u . $ Here, two steps are used:
\begin{enumerate}
\item Assume a transform, $ \phi(u) , $
having a specific parametric form in $ u , $
and require that $ \phi $ be governed by a simple
(linear, non-advective) diffusion equation.
The form of $ \phi $ used is given by equation \eqref{transform}.
\item Insert the assumed $ \phi(u) $ into the
diffusion equation and compare the result (equation \eqref{cole1})
with the original nonlinear governing equation (\eqref{burgers} below);
attempt to determine a detailed form for $ \phi(u) $
by forcing the former equation to match the latter.
\end{enumerate}
\end{enumerate}
Here, and
in response to \cite{garba} who
noted the arbitrary origin of the Cole-Hopf solution to Burger's equation,
we find a suitable $ \phi(u) $ and thus,
a fundamental
derivation of the Cole-Hopf solution.

Burger's equation, 
\begin{equation}\label{burgers} 
u_{t} + u u_x - \nu u_{xx} = 0
\end{equation}
which gained prominence as a simplified model
of compressible irrotational flow \cite{burgers}, has since
found application in a wide range of other problems,
including shock dynamics \cite{whitham},
nonlinear acoustics \cite{nimmo}, magnetohydrodynamics \cite{galtier}, 
turbulence \cite{gurarie}, traffic flow \cite{aw}, dynamics
of dislocations, polymer chains, and vortex lines \cite{ertas}, and 
formation of large scale cosmic structure
\cite{vergassola}. 
Variants of Burger's equation also appear in models of
flame front propagation and forced diffusion \cite{medina}.

In the following, the source-free Kardar-Parisi-Zhang (KPZ) equation 
\begin{equation}\label{kpzeqn}
h_t + \frac{1}{2} h_x^2  - \nu h_{xx} = 0
\end{equation}
appears as a useful analytical intermediary. In brief, this equation
serves as a well-studied model of
interfacial growth \cite{kardar}, where
the local rate of change in interface height, $ h_t(x,t) , $ 
reflects surface smoothing, $ \nu h_{xx} , $ due to, e.g., 
condensation and/or evaporation, combined with
growth, $ h_x^2 , $ in the local surface-normal direction.
Here, we only require the well-known transformation between
the KPZ and Burger's equations:
\begin{equation}\label{kpzburgers}
u=h_x
\end{equation}
(where in multiple dimensions $ \mathbf{u} = \boldsymbol{\nabla} h ); $
see, e.g., \cite{fogedby}. In this case, Burger's equation describes
the nonlinear evolution of the interfacial slope.

%%%%%%%%%%%%%%%%%%%%%%%%%%%%%%%%%%%%%%%%%%%%%%%%%%%%%%%5
\subsection{Time step conditions allowing construction of 
approximate incremental $ G $
for backward-form adjoint problems}\label{analyticgreens}
The adjoint equation in $ G $ is, as shown below, 
of backward form, and thus proves difficult to solve.
However, by restating the adjoint equation in forward time
form and by properly choosing the time step size,
$ \Delta s , $ we can 
obtain an approximate analytical solution. This section presents a 
straightforward argument for determining appropriate
conditions on $ \Delta s .$ The conditions obtained are
general, applicable to any evolution problem 
in which advection and diffusion are extant.

To begin the incremental solution over $ t' \in (t_j,t_{j+1}] , $
replace both $ \eta (x',t') $ and $ -b(x',t') = - \mathbf{b} \cdot \mathbf{e_x} 
= u(x',t') $
in (\ref{proc1})-(\ref{moper}) with
$ u (x',t') , $ and set $ f = g = 0 . $
The associated adjoint equation, from (\ref{proc1adj}) and (\ref{gammaeqn}),
then assumes the form:
\begin{equation}\label{chapkol} 
G_{t'} + u(x',t') G_{x'} + G u_{x'} (x',t') +
\nu G_{x'x'} = 0  
\end{equation}
with
\begin{equation}\label{ginitcond}
G( x,t_{j+1}|x',t') \rightarrow \delta (x' -x) \ \ \mathrm{as} \ \  t' \rightarrow t_{j+1}
\end{equation}
where again $ t_j < t' \le t_{j+1} . $ 

Restate (\ref{chapkol}) 
(which is in backward form in the
forward time $ t' ) $
in forward form, in the backward time, $ s' : $
\begin{equation}\label{chapkol1} 
G_{s'} + b(x',t') G_{x'} + G b_{x'}(x',t') -
\nu G_{x'x'} = 0  
\end{equation}
with
\begin{equation}\label{ginitcond1}
G( x,s_{k}|x',s') \rightarrow \delta (x' -x) \ \ \mathrm{as} \ \  s' \rightarrow s_{k}
\end{equation}
where $ s_k \le s' < s_{k+1} , $ and where $ s_k $ and $ s_{k+1} $
correspond respectively to $ t_{j+1} $ and $ t_j . $ Note, $ \Delta s = s_{k+1} - s_k $
and $ \Delta t = t_{j+1}-t_j . $

Next, determine conditions on $ \Delta s $ that, first,
allow $ b(x',s') $ in (\ref{chapkol1}) 
to be approximated by $ b(x,s_k) , $
and second, allow neglect of the term
$ G b_{x'} ; $ given these conditions, (\ref{chapkol1}) can be linearized and
an analytical solution obtained. 

We proceed
heuristically. First note that since $ G $ satisfies
the same (Fokker-Planck) equation and initial condition as
an incremental transition density function,
$ p(x,s_k | x', s') , $ associated with a stochastic 
process
\begin{equation}\label{incchi}
d \chi(s') = b(\chi(s'),s') (s'-s_k) + \sqrt{2 \nu} [w(s')-w(s_k)]
\end{equation}
then for small $ \Delta s , $
the random walk swarms governed by
(\ref{incchi}) and launched from the $ x-\mathrm{axis} $ at $ s'= s_{k+1} ,$
toward the backward time slice $ s' = s_{k} $ (i.e., the current forward time, $ t ) : $
i) will at $ s_{k+1} , $ have an approximate mean launch position
$ \bar{x}(s_{k+1}) \approx x + b(x,s_k) \Delta s , $
and ii) will
have non-negligible probability of reaching the solution point $ ( x, s_k ) $ only
if they lie within an approximate distance
$ \sqrt{\nu \Delta s } $
of $ \bar{x}(s_{k+1}) . $
Hence, 
the size of
the region over which both $ G = p $ and
gradients in $ G $ are non-negligible, is on the order
of $ \sqrt{\nu \Delta s } . $
 
Denoting the length and time scales
associated with the drift field $ b $ as $x_s$ and $t_s,$
respectively, and noting that the respective scales of the four terms
in (\ref{chapkol1}) are $ O(G_s/\Delta s) , $
$ O(b_s G_s/\sqrt{\nu \Delta s}) , $ $ O(b_s G_s/x_s) , $
and $ O(G_s/\Delta s) , $ where $ G_s $ and $ b_s $  
are the local Green's function and drift scales, then we find that
the first condition on $ \Delta s , $ allowing neglect of the
$ G b_{x'} $ in (\ref{chapkol1}), is
\begin{equation}\label{timestepcond1}
\frac{\sqrt{\nu \Delta s}}{x_s} << 1
\end{equation}

A second condition,
\begin{equation}\label{timestepcond2}
\frac{\Delta s}{t_s} << 1
\end{equation}
allowing replacement of $ b(x',s') $ with $ b(x,s_k) ,$
follows by expanding $ b(x',s') $
about $ (x,s_k) :$ 
\begin{displaymath}
b(x',s')=b(x,s_k) + O \big[ \frac{\sqrt{\nu \Delta s}}{x_s}, \frac{\Delta s}{t_s} \big]
\end{displaymath}
where $ x'-x = O(\sqrt{\nu \Delta s}) $
and $ s'-s = O(\Delta s) . $

Assuming that $\Delta s $ is chosen so that 
(\ref{timestepcond1}) and (\ref{timestepcond2}) hold,
then (\ref{chapkol1}) can be solved via Fourier
transform
\begin{equation}\label{greenfourier}
\hat{G} (x , s_k | k , s') = \int_{-\infty}^{\infty} e^{-2 \pi i k x'}
G(x,s_k| x',s') dx'
\end{equation}
with the result
\begin{equation}\label{incgreensoln}
G(x,s_k|x',s') = \frac{1}{\sqrt{4 \pi \nu (s' - s_k)}}
\exp \Big[ \frac{-\big[x' - \big(x + b(x,s_k)(s'-s_k)\big)]^2}{4 \nu (s'-s_k)} \Big]
\end{equation}
This can be restated in terms of physical (forward) time
variables using $ s'-s_k = t_{j+1}-t' $ and $ b(x,s_k) = -u(x,t_{j+1}) =
-u(x,t_j) + O(\Delta t) , $ to arrive at
\begin{equation}\label{incgreensoln1}
G(x,t_{j+1}|x',t') = \frac{1}{\sqrt{4 \pi \nu (t_{j+1}-t')}}
\exp \Big[ \frac{-\big[x' - \big(x - u(x,t_j)(t_{j+1}-t')\big)]^2}{4 \nu (t_{j+1}-t')} \Big]
\end{equation}

Thus, using (\ref{incgreensoln1}) in (\ref{diffgreensoln}) (with $ f $ again set equal to
$ 0), $ we obtain the
incremental Green's function solution to Burger's equation:
\begin{equation}\label{diffburgers}
u(x , t_{j+1}) = \int_{-\infty}^{\infty} u (x', t_{j} ) 
\Big[ \sqrt{4 \nu \pi (t_{j+1} - t_j)} \Big]^{-1} 
\exp -[(x'- \bar{x}_{j+1})^2/(4 \nu (t_{j+1}-t_j) )] dx'
\end{equation}
where $ \bar{x}_{j+1} = \bar{x}(t_{j+1}) = x - u(x,t_j) (t_{j+1}-t_j) .$ 
Since the incremental Green's function, $ G(x,t_{j+1}|x',t') $
corresponds to
an incremental transition density, $ p(x,t_{j+1}|x',t') , $
then this solution can also be interpreted 
probabilistically as an explicit version of the 
representative solution in (\ref{diffsol1a}).
%%%%%%%%%%%%%%%%%%%%%%%%%%%%%%%%%%%%%%%%%%%%%%%%%%%%%%%%%%%%
\subsection{Derivation of the Cole-Hopf solution}\label{linnonlin}
Although the presence of the
term $ u(x, t_j) $ in the exponential in (\ref{diffburgers})
does not prevent, e.g., numerically-based solutions,
in order to both validate the time-incremental approach
as well as explore potential approaches for
analytically tackling other nonlinear evolution problems,
we now focus on using (\ref{diffburgers}) to obtain the non-incremental
Cole-Hopf solution. 

The path connecting (\ref{diffburgers})
to the Cole-Hopf solution is indicated by the \textit{form}
of the incremental solution. In particular, we recognize that finding
a non-incremental solution requires, at minimum,
elimination of the \textit{a priori} unknown drift, $ u(x',t') , $
from the incremental Green's function in (\ref{incgreensoln1}); elimination of 
the drift
is required in order to 
stretch the time increment, $ \Delta t' , $ to arbitrary
lengths. 

Thus, we 
introduce a surrogate, $ \phi = \phi(u) , $ for $ u $ 
upon which we impose two key requirements. First, 
in order to take advantage of the 
machinery developed above,
we require that $ \phi (u) $ also satisfies
an incremental evolution equation of the 
same form as (\ref{diffburgers}):
\begin{equation}\label{diffphi}
\phi(x, t_{j+1}) = \int_{-\infty}^{\infty} \phi(x', t_{j} ) 
G^{\phi}(x,t_{j+1}| x',t_j) dx'
\end{equation}
where $ G^{\phi} $ is the incremental Green's function
associated with the evolution of $ \phi . $
Second, 
in order to eliminate $ u $ from the incremental
Green's function, $ G^{\phi} , $
we require that the evolution of $ \phi $
be purely diffusive:
\begin{equation}\label{govphi}
\phi_{t'} - \nu \phi_{x'x'} = 0 
\end{equation}
so that the associated adjoint equation is
\begin{equation}\label{gammaphi}
G_{t'}^{\phi} + \nu G_{x'x'}^{\phi} = -\delta(t'-t) \delta(x'-x) 
\end{equation}

The Green's function, $ G^{\phi} , $
is easily obtained by setting $ u = 0 $
in (\ref{incgreensoln1}); 
the detailed 
incremental 
solution in (\ref{diffphi}) then follows:
\begin{equation}\label{phieqn}
\phi(x, t_{j+1}) = \int_{-\infty}^{\infty} \phi(x', t_{j} ) 
\Big[ \sqrt{4 \pi \nu (t_{j+1} - t_j)} \Big]^{-1} 
\exp -[(x'- x)^2/(4 \nu (t_{j+1}-t_j) )] dx'
\end{equation}
where now $ \bar{x} = x . $

In order to proceed, we next recognize that by assuming a specific
parametric form for $ \phi (u) , $
followed by introduction of this guessed form into \eqref{govphi},
an evolution equation in $ u $ emerges. If the latter can be transformed
or forced to match the actual equation for $ u ,$ 
\eqref{burgers}, then we will have constructed a transform from
the incremental to the non-incremental solution. 
Thus, after a few trials, we choose
\begin{equation}\label{transform}
\phi (u) = \exp ( f(u) ) 
\end{equation}

Inserting (\ref{transform}) into
(\ref{govphi}) then yields
\begin{equation}\label{cole1}
u_{t'} - \nu u_{x'x'} - \nu [ f' + \frac{f''}{f'}] u_x^2 = 0
\end{equation}
where 
$ f' = df/du . $

No choice of $ f(u) $ allows
(\ref{cole1}) to be placed in the form of Burger's equation
(\ref{burgers}); however, by redefining
$ u $ as $ h , $ and choosing
\begin{equation}\label{subs}
f' + \frac{f''}{f'} = - \frac{1}{2 \nu}
\end{equation}
we observe that (\ref{cole1}) transforms to the KPZ equation (\ref{kpzeqn}).
Replacing $ u $ with $ h $ in (\ref{cole1}) and differentiating 
the result with respect to $ x' $ then yields
\begin{equation}\label{cole2}
v_{t'} - \nu v_{x'x'} - 2 \nu v v_{x'} [ f' + \frac{f''}{f'}] - \nu v^3 
[f' + \frac{f''}{f'}]' = 0
\end{equation}
where, for clarity, we express $ h_{x'} $ as $ v , $ and where 
the last bracketed term is differentiated
with respect to $ u . $
Thus, by (\ref{subs}),
(\ref{cole2}) transforms to Burger's equation (\ref{burgers}),
and as noted, (\ref{cole1}) transforms to the KPZ equation (\ref{kpzeqn}).

Although a number of solutions for $ f(u) $ via (\ref{subs})
are available, we choose $ f' = - 1 / (2 \nu ) $
[where it is understood that $ u $ in (\ref{transform}) now corresponds to $ h $ in
(\ref{kpzeqn}) and $ v $
in (\ref{cole2}) corresponds to $ u $ in (\ref{burgers})].
Thus, from (\ref{transform})
\begin{equation}\label{phidetail}
\phi(h)= \phi(x',t')= \exp{\big( \frac{ - h}{2 \nu} \big)}
\end{equation}
or inverting,
\begin{equation}\label{hdetail}
h (x',t') = - 2 \nu ln ( \phi(x',t') )
\end{equation}
which corresponds to the Cole-Hopf transformation for the
KPZ equation \cite{fogedby}.

The corresponding Cole-Hopf transform for Burger's equation,
which we denote as $ \tilde{\phi}(v) , $
then follows from (\ref{phidetail})
via integration of $ v = h_x : $ 
\begin{equation}\label{phiburgers}
\tilde{\phi}(v)= \tilde{\phi}(x',t')= \exp{\big( \frac{ - \int_0^{x'} v(x'',t') dx'' }{2 \nu} \big)}
\end{equation}

Returning to the incremental solution (\ref{diffphi}) for $ \tilde{\phi} $
(with $ \phi $ is restated as $ \tilde{\phi} ) , $
we insert (\ref{phiburgers}) and follow Whitham \cite{whitham}
by isolating 
$ v(x, t_{j+1}) : $ 
\begin{equation}\label{uisolate} 
v (x , t_{j+1}) = -2 \nu \Big[ \frac{\phi_x}{\phi} \Big]_{x,t_{j+1}}  
\end{equation}
or
\begin{equation}\label{uisolate1}
v(x,t_{j+1}) =  2 \nu
\frac{ \Big[ \sqrt{4 \pi \nu (t_{j+1} - t_j)} \Big]^{-1} 
\int_{-\infty}^{\infty} \phi(x', t_{j} ) 
\exp -[(x'- x)^2/(4 \nu (t_{j+1}-t_j) )] \frac{(x'-x)}{2 \nu (t_{j+1}-t_j)} dx'}
{\Big[ \sqrt{4 \pi \nu (t_{j+1} - t_j)} \Big]^{-1}
\int_{-\infty}^{\infty} \phi(x', t_{j} ) 
\exp -[(x'- x)^2/(4 \nu (t_{j+1}-t_j) )] dx'}
\end{equation}
Finally, set $ t_{j+1} = t , $ $ t_{j} = 0 , $ and
after some manipulation, obtain
the classic Cole-Hopf solution of Burger's equation (see, e.g., \cite{whitham} ):
\begin{equation}\label{burgersfin}
v(x , t) = \frac{ \int_{-\infty}^{\infty} \frac{(x-x')}{t} \exp (-H/(2 \nu)) dx'}
{ \int_{-\infty}^{\infty} \exp (-H/(2\nu)) dx'}
\end{equation}
where $ H = H (x';,x, t) = \int_0^{x'} v(y,t=0) dy + \frac{(x-x')^2}{2t} . $
%%%%%%%%%%%%%%%%%%%%%%%%%%%%%%%%%%%%%%%%%%%%%%%%%%%%
\section{Test Case 2: Soliton solution of
the nonlinear Schr\"odinger equation}\label{nonlinex2}
As a second test,
we consider solution
of the one-dimensional nonlinear Schr\"odinger equation
\begin{equation}\label{cubicshrodinger}
i \eta_{t'} + \eta_{x' x'} + \kappa |\eta|^2 \eta = 0
\end{equation}
Like Burger's equation, (\ref{cubicshrodinger}) represents a canonical nonlinear
evolution equation \cite{whitham}, capturing in this instance
nonlinearly
dispersive, weakly dissipative wave propagation. As with Burger's equation, 
(\ref{cubicshrodinger}) and its variants appear in a wide range of
contexts, including nonlinear optics \cite{whitham}, 
hydrodynamics \cite{zakharov}, and
plasma physics \cite{kateskaup}.

Here, we wish to show how a well-known soliton solution
to (\ref{cubicshrodinger}) can be obtained using the 
machinery developed above. In contrast to the
function transform approach used in Test Case 1, however,
we make the jump from incremental
to non-incremental solution via asymptotics;
such approaches can be considered, e.g., 
when nonlinear terms in the governing evolution equation are,
on the scales of interest, small.

To begin, we seek a 
traveling wave solution \cite{whitham} of the form:
\begin{equation}\label{nlseguess}
\eta (x',t') = \exp (i gx' - i ft') h(X') 
\end{equation}
where $ X' = x' - U t' $ is a coordinate attached
to the moving wave, $ U $ is the wave speed, and $ g $ and $ f $ are constants.

We focus on determining the wave envelope shape, $ h (X') , $ and simply 
note that $ g $ and $ f $ can be determined in terms
of $ U $ using, e.g., Whitham's approach \cite{whitham}.
Thus, (\ref{cubicshrodinger}) is re-expressed in the wave-fixed
coordinate system as
\begin{equation}\label{nlse2}
i \eta_{t'}  - U \eta_{X'} + \eta_{X'X'} + \kappa | \eta|^2 \eta = 0
\end{equation}
The corresponding Green's function, again valid over short
time intervals, $ \Delta t' , $ is governed by
\begin{equation}\label{greennlse}
- i G_{t'}  - U G_{X'} + G_{X'X'} + \kappa | \eta|^2 \eta = -\delta(X-X') \delta(t-t')
\end{equation}

In order to determine $ G , $ we 
take the spatial Fourier transform of (\ref{greennlse}),
\begin{displaymath}
\hat{G} = \int_{-\infty}^{\infty} \exp (-2 \pi i k X' ) G(x,t|X',t') dX'
\end{displaymath}
to obtain:
\begin{equation}\label{greenhat}
G(X,t|X',t')= \frac{1}{2\pi} \int_{-\infty}^{\infty} \exp[2 \pi i k (X'-X)]
\exp \big[ [-2 \pi U k + i 4 \pi^2 k^2] (t'-t) \big] dk
\end{equation}

The incremental solution over $ \Delta t' $ then follows from
(\ref{diffgreensoln}):
\begin{eqnarray}\label{diffnlse}
\eta ( X , t ) = & \frac{ e^{i(gx -ft)}}{2 \pi} \int_{- \infty}^{\infty} 
\int_{t}^{t+\Delta t} \int_{-\infty}^{\infty} h^3 (X') e^{2 \pi i U t} 
\exp \big[ [-2 \pi U k + i 4 \pi^2 k^2] \tau' \big] dk dX' d \tau' + \nonumber \\
 & + \int_{-\infty}^{\infty} \int_{-\infty}^{\infty} \frac{1}{2 \pi} 
e^{2 \pi i k (\Delta X') } 
\exp \big[ [-2 \pi U k + i 4 \pi^2 k^2] \Delta t' \big] 
h(X') e^{g X' - f (t + \Delta t')} dk dX' 
\end{eqnarray}
where $ \tau' = t' - t $ and $ \Delta X' = X' - X . $ 

Focusing first on the first term on the right of (\ref{diffnlse}) and carrying out the
time integral,
we note that since $ (2 \pi)^{-1} \int_{-\infty}^{\infty} e^{2 \pi i k \Delta X'} =
\delta ( \Delta X') = \delta [ X' - (x - U_o t) ] , $ then
this term becomes $ e^{i(gx - ft)} h^3(X) \Delta \tau , $ where, for clarity,
we write $ U_o = -U , $ and where $ \Delta \tau = \Delta t' . $

Turning to the second term in (\ref{diffnlse}) and using similar
steps, we arrive at
\begin{eqnarray}\label{term2}
\int_{-\infty}^{\infty} \int_{-\infty}^{\infty} \frac{1}{2 \pi} 
e^{2 \pi i k \Delta X'} h (\tilde{X}') 
\exp \Big[ i \big[g \big(\tilde{X}' + U_o(t + \Delta \tau) \big) -
f ( t + \Delta \tau ) \big] \Big] d \tilde{X}' dk = &  \nonumber \\
\int_{-\infty}^{\infty} \delta (\Delta \tilde{X}') h (\tilde{X}')
\exp \Big[ i \big[g \big(\tilde{X}' + U_o(t + \Delta \tau) \big) d \tilde{X}' 
h \big(x - U_o(t+\Delta \tau) \big) \exp \big[ i \big(g x - f (t + \Delta \tau) \big) \big]
\end{eqnarray}
where $ \Delta \tilde{X}' = \tilde{X}' - ( x - U_o (t + \Delta \tau)) , $
and where the term $ \exp [2 \pi U_o k + i 4 \pi^2 k^2 ] \Delta \tau $
has been replaced by $ 1 + O(\Delta \tau ) . $

The incremental solution (\ref{diffnlse}) thus assumes
the form
\begin{equation}\label{nlsesoln}
h(x - U_o t) e^{i(gx - ft)} = \kappa \Delta \tau h^3 (x - U_o t) e^{i(gx-ft)}
+h (x - U_o (t + \Delta \tau)) e^{i(gx-ft)} e^{if \Delta \tau}
\end{equation}
Canceling the common exponential term and expanding the
last term above about $ X = x - U_o t , $ then yields
\begin{equation}\label{nlsesoln1}
h(X) = \kappa \Delta \tau h^3 (X) + e^{if \Delta \tau} \big[
h(X) - h_{X} \Delta X + \frac{1}{2} h_{XX} \Delta X^2 \big] + O(\delta X^2)
\end{equation} 
or equivalently,
\begin{equation}\label{nlsesoln2}
\frac{f^2 \Delta \tau^2}{2} h(X) = \kappa \Delta \tau h^3 (X) + \frac{1}{2} h_{XX} \Delta X^2
- h_X \Delta X
\end{equation} 
where $ \Delta X = U_o \Delta \tau . $ 

In the last expression, we observe a separation of scales, i.e., the term
involving $ h_X $ is of $ O (\Delta \tau) , $
while all remaining terms are either of order $ \Delta \tau^2 $ or $ O(\kappa \Delta \tau ) . $
This becomes clear by expressing (\ref{nlsesoln2}) in dimensionless form
\begin{equation}\label{nlsesoln3}
\tilde{h}_{\tilde{X}\tilde{X}} \Delta \tilde{X}^2 + \tilde{\kappa} \tilde{h}^3 - 
\tilde{\beta} \tilde{h} - \tilde{h}_{\tilde{X}} \Delta \tilde{X}
\end{equation} 
where $ \tilde{h}=h/h_s , $ $ \tilde{X} = X/x_s , $ $ \Delta \tilde{X} =
\Delta X / \Delta x_s , $ $ \tilde{\kappa} = 2 \kappa \Delta \tau
h_s^2 x_s^2 /\Delta x_s^2 , $ $ \tilde{\beta} = f^2 \delta \tau^2 x_s^2 / \Delta x_s^2 , $
$ \Delta x_s = U_o \Delta \tau , $ and where $ h_s $ and $ x_s $ denote
the amplitude and axial length scale of the 
wave envelope, $ h(X) . $
[Note, $ \Delta \tilde{X} $ is $ O(1) .] $
 
Thus, expressing $ \tilde{h} $ for example, as 
\begin{equation}\label{perturbsoliton}
\tilde{h} = \tilde{h}_o + \epsilon
\tilde{h}_1 + O(\epsilon^2) 
\end{equation}
where $ \epsilon = \Delta x_s / x_s , $
and focusing on the case where
$ \tilde{\kappa} $ and $ \tilde{\beta} $ are both positive and $ O(1) ,$
we obtain a physically and mathematically consistent
solution for $ h(X) : $ 
\begin{equation}
\tilde{h}_o = 0
\end{equation}
and
\begin{equation}\label{solitonsoln}
\tilde{h}_1 = \big( \frac{2 \tilde{\beta}}{\tilde{\kappa}} \big)^{1/2}
sech \big( \frac{\tilde{\beta}}{\Delta \tilde{X}^2} \big)^{1/2} \tilde{X}
\end{equation}
which is a well-known soliton solution \cite{whitham}
of the cubic Schr\"odinger equation (\ref{cubicshrodinger}). 

As a closing summary to this and Test Case 1, 
we have shown that a systematic attack on nonlinear
evolution problems can be initiated from
an incremental Green's function solution. In most problems,
one would typically proceed to a numerical
time integration, using the incremental GF 
to construct the kernel. Here, for purposes of validation,
analytical integration
has been pursued.
%%%%%%%%%%%%%%%%%%%%%%%%%%%%%%%%%%%
\section{The GFSM as a physical probe: 
organization of near-molecular-scale vorticity in Burger's 
vortex sheets}\label{burgersvortsec}
The last example studies physical features underlying evolution of
single, multiple, and continuous sets of Burger's vortex
sheets evolving within deterministic and random strain rate
fields.
The example is designed to illustrate application of Green's
function and stochastic process ideas as 
probes for exploring linear and nonlinear evolution problems.
Since the example is long, encompassing three distinct elements,
we expand upon the introductory
overview.

The generic problem of 
evolution of $ N_x $ Burger's vortex sheets (BVS), generated by the  
appearance of an arbitrary, spatially discrete, one-dimensional
velocity field is first presented (section \ref{sec6.1}).
A discrete initial velocity condition is chosen
since it leads to a physically and mathematically crucial
delta function initial condition on the vorticity evolution problem.

Mathematically, as detailed in section 6.2, this condition
allows calculation of an essential single sheet Green's function;
the importance of this GF emerges when it is reinterpreted
as a transition density and applied to compute
statistical behavior of single, multiple, and continuous
sets of BVS's evolving within random strain rate fields
(sections 6.5-6.7).  
 
Physically, and as mentioned, the delta function IC allows
us to interpret the vorticity transport equation as a Fokker-Planck
equation for an underlying OU process (section 6.3).
Presuming that the
OU process describes the stochastic dynamics
of a sub-BVS physical entity, viz, 
elemental vortex sheets,
we are led to investigate vorticity dynamics
within individual Burger's sheets (section \ref{sec6.3}).
This examination leads to the following observations and results:
\renewcommand{\labelenumi}{\arabic{enumi})}
\begin{enumerate}
\item On short acoustic time and near-molecular-length-scales, scaling shows that
in-sheet vorticity is disordered, three-dimensional, and diffusive. 

\item Since vorticity, on the long BVS time-scale, becomes one-dimensional and
highly organized,
some organizing mechanism clearly operates over the
longer time scale. 

\item Presuming that organization is effected  
by weak, in-sheet hydrodynamic modes, 
we investigate
these modes using a simple analog: sub-sheet vorticity organization in
\textit{unstrained}
planar vortex sheets. 

\item Finally, the modal analysis suggests that 
three hydrodynamic mechanisms underlie in-sheet organization:
\begin{enumerate}
\item weakly-damped, cross-sheet acoustic modes,

\item a diffusive cross-sheet shear mode, and 

\item a diffusive cross-sheet entropy mode.
\end{enumerate}
\end{enumerate}

Burger's vortices were first studied by Burgers in 1948 \cite{burgers} 
who showed that such structures are capable of dissipating turbulent
kinetic energy through the combined action of viscous dissipation and 
vortex stretching. Soon after, Townsend \cite{townsend}
proposed that the fine structure of high Reynolds number
turbulence, i.e., the structure extant on scales smaller than the 
viscous dissipation scale, is characterized by random distributions of 
Burger's line and sheet vortices. Since these early works,
Burger's vortex lines and sheets have been studied both as
a fundamental, analytically tractable model of the combined action
of viscosity, advection, and stretching on vorticity,
see, e.g., \cite{saffman,sherman}, and as a putative fundamental structure
in various turbulent flows \cite{pullin}.   
%%%%%%%%%%%%%%%%%%%%%%%%%%%%%%%%%%%%%%%%%%%%%%%%%%%%%%%%%%%%
\subsection{Initial conditions and governing equation}\label{sec6.1}
Attention is limited to the case where
a series of parallel vortex sheets are formed at some instant, $ t = t_o , $
within a two dimensional potential flow, i.e., a strain rate field, given
by $ \mathbf{u} = [-kx , 0 , kz] . $  See figure 1.
The formation of each sheet occurs due to the appearance of a spatially varying
flow component in the y-direction, given by 
\begin{equation}\label{vcomp}
v (x,t=t_o) =  \sum_{i=1}^{N_x} \Delta v_i U(x-x_i)
\end{equation}
where $ v = v_y , $ $ \Delta v_i $ is the incremental velocity at
$ x_i , $ $ U (x - x_i) $ is the unit step function, 
and $ N_x $ is the number of increments. 
There is no limit on the size of individual velocity
increments; as shown immediately below, these determine the strength of each
associated vortex sheet.

\begin{figure}
\centering
\includegraphics[width=5.0in]{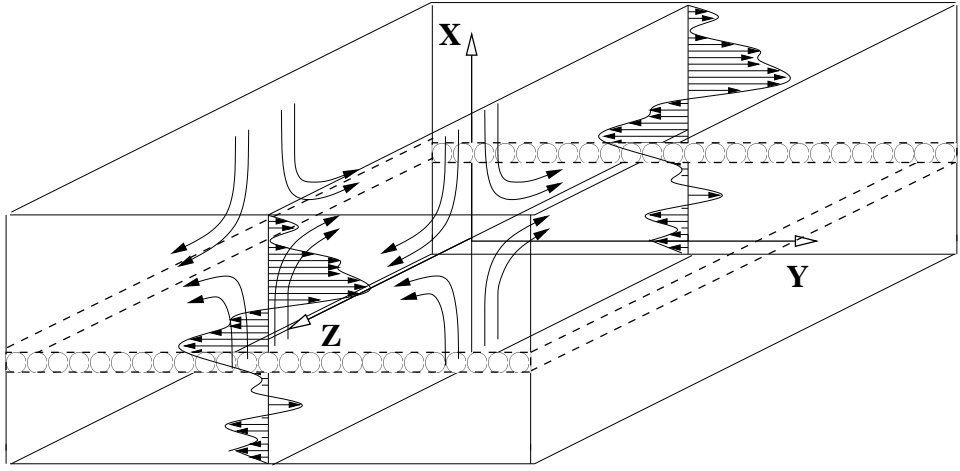}
\caption{Continuous collection of Burger's vortex sheets evolving
under influence of an inviscid strain rate field, $ v_x = -kx , v_z = k z . $
Vorticity is one-dimensional
in the $ z-\mathrm{direction} $ and is produced by
the $y-\mathrm{velocity} $ component, $ v_y (x,t) . $ }
\end{figure}

The vorticity equation assumes the linear 1-D form
\begin{equation}\label{burgervortexgoveqn}
\Omega_{t} - kx \Omega_{x} = k \Omega + \nu \Omega_{xx}
\end{equation}
where $ \Omega $ is directed along the $z-\mathrm{axis.} $
The initial condition associated with (\ref{vcomp}) is
\begin{equation}\label{vortexic}
\Omega(x,t=t_o) = \sum_{i=1}^{N_x} \Delta v_i \delta (x-x_i)
\end{equation}
Due to the linearity of the governing equation (\ref{burgervortexgoveqn}),
the last condition can be given a clear
physical meaning: it corresponds to the superposition of
$ N_x , $ infinitesimally thin vortex sheets, $ \{ \Omega_i \}_1^{N_x} , $
\begin{equation}\label{vortexic1}
\Omega(x,t=t_o) = \sum_{i=1}^{N_x} \Omega_i (x_i,t_o) 
\end{equation}
where
\begin{equation}\label{ivortexic}
\Omega_i (x_i,t_o) = \Delta v_i \delta (x-x_i) 
\end{equation}
(and where units on $ \delta (x-x_i) = \frac{dU}{dx} $ 
are $ \mathrm{length}^{-1} ).$

The generic solution, describing the vorticity field
produced by the $ N_x $ sheets, is of the form:
\begin{equation}\label{vortexgenlsoln}
\Omega(x,t) = \sum_{i=1}^{N_x} \Omega_i (x,t) 
\end{equation}
where the detailed form of $ \Omega_i $ is given below.
%%%%%%%%%%%%%%%%%%%%%%%%%%%%%%%%%%%%%%%%%%%%%%%%%% 
\subsection{Derivation of the single sheet Green's function}\label{sec6.4}
We derive the single sheet Green's function,
$ G^{(i)}(x,t|x_i,t_o) , $
by first normalizing \eqref{burgervortexgoveqn} with the initial
sheet strength, $ \Delta v_i , $ and then by noting that 
the resulting problem takes the form 
of that governing a forward time GF; see, e.g., \cite{barton}.
Thus, we solve for $ \Omega_i(x,t) / \Delta v_i , $
and use 
\begin{equation}\label{gfomega}
G^{(i)} (x,t|x_o,t_o) = \frac{\Omega_i (x,t)}{\Delta v_i}
\end{equation}
where $ x_o $ is the initial location of the infinitely thin vortex sheet.
We take $ t_o = 0 , $ temporarily drop 
superscripts and subscripts referring to sheet $ i , $ and
refer to the vorticity field produced as an individual
Burger's vortex sheet.  

While Townsend's 1951 paper
\cite{townsend} is cited by Saffman \cite{saffman} as the source
for a solution to a scalar transport problem 
occurring in a time-varying strain field of the type considered here,
Townsend's derivation, based on the method of characteristics (MOC),
is apparently given elsewhere.
Here, using a Fourier transform-MOC approach,
representing a slightly generalized version
of Gardiner's solution for a one-dimensional OU
process evolving in a steady drift field \cite{gardiner},
we first derive a solution
appropriate to time-varying strain fields of the form:
\begin{equation}\label{backgroundflow}
\mathbf{v} = -k(t)x \mathbf{e_x} + k(t) z \mathbf{e_z}
\end{equation}

Fourier transforming (\ref{burgervortexgoveqn}) yields
\begin{equation}\label{fouriervortex}
\hat{\Omega}_t + k(t) \kappa \hat{\Omega}_{\kappa} 
+ \nu \kappa^2 \hat{\Omega}
= 0
\end{equation}
Placing this in characteristic form then yields:
\begin{equation}\label{character1}
\frac{ d \hat{\Omega}}{d t} = - \nu \kappa^2 \hat{\Omega}  \hspace{.7cm} \mathrm{on}
\hspace{.7cm} \frac{d \kappa}{d t} = k(t) \kappa
\end{equation}
Integrating along characteristics from $ \kappa_o = \kappa (t_o) $ to 
$ \kappa = \kappa(t) $
yields 
\begin{equation}\label{seqn}
\kappa(t)= \kappa_o \exp [h(t)]
\end{equation}
where
\begin{equation}\label{heqn}
h(t) = \int_{t_o}^t k(t') dt'
\end{equation}

Next, substituting (\ref{seqn}) in (\ref{character1}) gives
\begin{equation}\label{omegaseqn}
\hat{\Omega}(\kappa, t) = \hat{\Omega} (k_o, t_o) \exp \big[ -\nu k_o^2 
\int_{t_o}^t e^{2 h(t')} dt' \big]
\end{equation}
while setting the initial time $ t_o = 0 $ and using the 
initial condition 
\begin{equation}\label{icvortex}
\Omega (x,0) = \Delta v
\delta(x-x_o)
\end{equation}
leads to 
\begin{equation}\label{icfouriervortex}
\hat{\Omega}(\kappa_o , 0) = \Delta v e^{i \kappa_o x_o } = \Delta v
\exp[ i \kappa e^{-h(t)} x_o ]
\end{equation}
Finally, letting
\begin{equation}\label{pdefn}
p(t) = e^{-2 h(t)} \int_0^t e^{2 h(t')} dt'
\end{equation}
and taking the inverse transform of (\ref{omegaseqn}) leads
to the final single-sheet Green's function:
\begin{equation}\label{finalvortexsoln}
G(x,t|x_o,t_o=0) = \frac{\Omega(x, t)}{\Delta v} = \frac{1}{\sqrt{ 4 \pi \nu p(t)}} 
\exp{ \big[ \frac{ -(x -x_o e^{-h(t)})^2}{4 \nu p(t)} \big]}
\end{equation}
The single sheet GF shows that peak response
to the initial delta function travels to $ x(t) = x_o e^{-h(t)} $
and spreads as $ p(t) . $
As a quick check, we note that in the case where $ k(t) $
is constant, (\ref{finalvortexsoln}) assumes the appropriate form
\cite{sherman, saffman}. 

In order to circumvent destabilizing vortex
compression, it appears that for most times, $ t , $
$ k(t) $ must remain positive. 
Although 
short periods of vortex compression, sandwiched between 
long periods of stretching, can likely be stably sustained, we make no attempt
to address this question. Rather, we limit attention, in the deterministic case, to 
$ k(t) > 0 , $ and
in the random case, 
$ k(t) = k_o + k'(t) , $
to positive $ k_o , $ with $ k_o > | k'(t)|_{max} , $
where $ k_o $ and $ k'(t) $ are the non-random and random
parts of $ k(t) . $
%%%%%%%%%%%%%%%%%%%%%%%%%%%%%%%%%%%%%%%%%%%%%%%5
\subsection{Vortex sheet evolution
as an Ornstein-Uhlenbeck process}\label{sec6.2}
Focusing on the evolution of an individual BVS, say the 
$ i^{th} $ sheet (having vorticity $ \Omega_i (t) )$
in a collection of $ N_x $ sheets,
again defining a normalized 
vorticity, $ \tilde{\Omega}_i = \Omega_i / \Delta v_i ,$
and again noting the initial condition \eqref{ivortexic},
we observe that equation 
(\ref{burgervortexgoveqn}) can also be interpreted as a Fokker-Planck equation
governing the transition density, 
\begin{equation}\label{pvortexeqality}
p^{(i)}(x,t|x_i,t_o) = \frac{\Omega_i(x,t)}{\Delta v_i}
\end{equation}
of an OU stochastic process $ \chi^{(i)} (t) , $
where $ \chi^{(i)} $ evolves as:
\begin{equation}\label{vortexchi}
d \chi^{(i)} (t) = -k (t) \chi^{(i)}(t) dt + \sqrt{2 \nu} d w^{(i)}(t)
\end{equation} 

Importantly, this connection allows us to introduce
a correspondence between the
evolution of individual realizations of the stochastic process, 
$ \chi^{(i)} (t) , $
and evolution of individual elemental vortex sheets. 
In this picture, 
the instantaneous BVS, here the $ i^{th} $
sheet, corresponds to a \textit{cloud}
of say $ N $ elemental vortex sheets; see figure 2. Although
we suppose for the moment that EVS's are
quasi-physical entities, analogous
to say idealized, infinitely thin vortex sheets, we argue below
that a physically reasonable embodiment of these can be defined;
see section \ref{sec6.3.2}.

\begin{figure}
\centering
\includegraphics[width=5.0in]{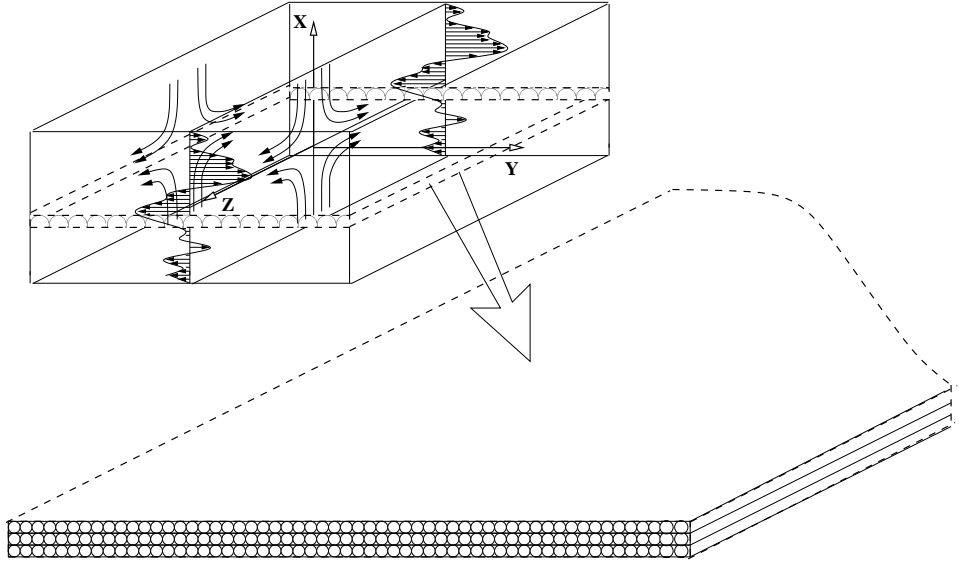}
\caption{Decomposition of individual Burger's vortex sheets into
elemental vortex sheets.}
\end{figure} 

Given a stochastic picture of BVS evolution, we
can now easily write down physically meaningful
expressions for the mean
position of the vortex cloud $(=i^{th} $ Burger's vortex sheet)
$ < \chi^{(i)} (t) >_{x} ,$
and it's diffusive spread, i.e., its thickness, 
$ < ( \chi^{(i)}(t) - < \chi^{(i)}(t) >_x )^2  >_x : $
\begin{equation}\label{vortexmeanpos}
< \chi^{(i)} (t) >_x = \int_{-\infty}^{\infty} x p^{(i)}(x,t|x_i,t_o) dx
\end{equation}
\begin{equation}\label{vortexsigma}
< (\chi^{(i)} (t) - <\chi^{(i)}(t)>_x)^2 >_x = \int_{-\infty}^{\infty} (x- 
< \chi^{(i)}(t) >_x )^2 p^{(i)}(x,t|x_i,t_o) dx
\end{equation}
where expectations over the stochastic process, $ \chi^{(i)}(t) , $
are written in the notationally convenient form, 
$ < \cdot >_x . $ 

When the strain rate $ k(t) $ is random
(see sections 6.6 and 6.7),
equations (\ref{burgervortexgoveqn}),
(\ref{ivortexic}), (\ref{vortexchi})-(\ref{vortexsigma}) 
apply to any realization of $ k(t) . $
Thus, 
following, e.g., \cite{ave1990, ave1992a, ave1994}, 
the average sheet position and spread over random $ k $ will be computed as:
\begin{equation}\label{vortexmeanposk}
\big< < \chi^{(i)} (t) >_x \big> = 
\big< \int_{-\infty}^{\infty} x p^{(i)}(x,t|x_i,t_o) dx \big>
\end{equation}
and
\begin{equation}\label{vortexsigmak}
\big< < (\chi^{(i)} (t) - <\chi^{(i)}(t)>_x)^2 >_x \big> = 
\big< \int_{-\infty}^{\infty} (x- 
< \chi^{(i)}(t) >_x )^2 p^{(i)}(x,t|x_i,t_o) dx \big>
\end{equation}
where  expectations taken with respect to $ k $ will be denoted as
$ < \cdot > . $
%%%%%%%%%%%%%%%%%%%%%%%%%%%%%%%%%%%%%%%%%%%%%%%%%%%%
\subsection{Elemental vortex sheets}\label{sec6.3}
In order to develop a physically and mathematically
consistent picture of EVS's,
we limit attention to
incompressible flows and consider 
vorticity transport
on length and time scales that are small relative to
those associated with BVS dynamics.

In order to identify an appropriate time scale,
we first appeal to the Biot-Savart
law:
\begin{equation}\label{biotsavart}
\mathbf{v_s}(\mathbf{x},t) = -\frac{1}{4 \pi} \int 
\frac{\mathbf{r} \mathbf{\times} \boldsymbol{\omega} (\mathbf{y}, t) }{r^3} 
d \mathbf{y}
\end{equation} 
where $ \mathbf{r} = \mathbf{x} - \mathbf{y} . $
This kinematic result, which gives the instantaneous solenoidal (incompressible)
velocity field, $ \mathbf{v_s} , $ in terms of an integral
over the vorticity field, $ \boldsymbol{\omega} , $
indicates that the local magnitude and direction of the former 
\textit{instantaneously} senses and responds
to the integrated effects of the latter. 
In reality, 
sensing and response are acoustically
mediated. 

Thus, we surmise that in order to expose
sub-sheet dynamics, we should focus on processes taking place
on acoustic time scales. This choice, in fact,
proves advantageous since the acoustic scale lies well separated from
both the much longer BVS time scale, $ k^{-1} , $
and the much shorter molecular collision scale, $ \tau_c , $
given below; as will be shown, substantial insight into
sub-sheet rotational dynamics emerges on this intermediate scale.

For the EVS length scale, we focus 
on lengths that are 
large relative to molecular scales, $ d_o , $
and small relative to
the characteristic BVS thickness, 
$ \delta_i (t) , $
where, depending on whether the strain rate field is
deterministic or not, 
$ \delta_i (t) $ is given, respectively,
by (\ref{vortexsigma}) or (\ref{vortexsigmak}). 

Thus, as a means of gaining a conceptual foothold, it proves
useful to focus on the acoustic time scale, near-molecular
length scale rotational dynamics and evolution of in-sheet \textit{clumps},
fluid particles comprised of a fixed number of
say $ N_{cl} $ molecules. Limiting attention
to liquids, and given a characteristic (effective) molecular diameter 
$ d_o , $ the characteristic 
clump size is $ \delta_{cl} \approx N_{cl}^{1/3} d_o . $

For a short period, $ \tau_{cl} , $ any given clump
remains nominally intact and experiences
an incremental rotation, the magnitude of which
corresponds to the average rotation of all constituent
molecules. Since clump dispersion
occurs due to thermal motion of constituent molecules,
$ \tau_{cl} \approx \delta_{cl} /v_T , $ where $ v_T = \sqrt{3kT/m} , $
the 
thermal speed, is on the order of the sound speed, $ a_o . $

Since the ratio of $ \tau_{cl} $ to the collision time scale,
$ \tau_c = O(d_o/v_T) , $ is $ O(N_{cl}^{1/3} ), $
choosing $ N_{cl} $ to be on the order of say, $ 10^3 , $
gives $ \tau_{cl} / \tau_c = O(10) . $
Hence, as depicted in figure 3, we envision
that an initially smooth collection of clumps
becomes 'bumpy', i.e., thermally roughened, over a time 
interval of $ O(\tau_{cl} ) ; $
likewise, as noted in the caption, individual clumps
remain largely intact. 

\begin{figure}
\centering
\includegraphics[width=2.5in]{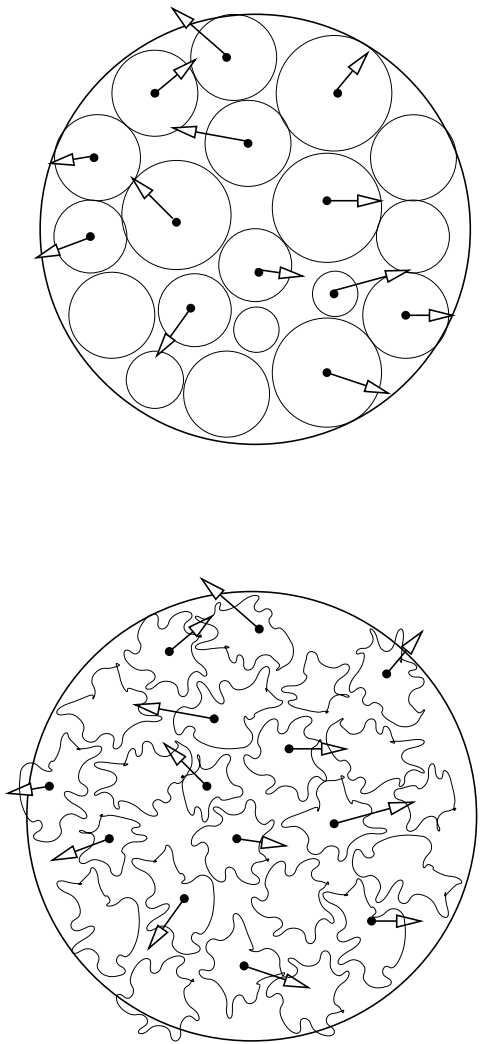}
\caption{Over a time interval on the order of
$ \tau_{cl} , $ an initially smooth
collection of neighboring clumps
begins to thermally disperse.  Since the clump
is large relative to individual constituent
molecules, dispersion on this time scale takes place predominantly via
thermal roughening and inter-penetration of surface molecules
(where the latter are depicted as surface bumps). Holes within,
and penetrating jets into, clumps
form as well, but the fraction of molecules
having sufficient energy to form these over $ \tau_{cl} $
is small; thus, these are not depicted.}
\end{figure} 

Indeed, the $ \tau_{cl}-\mathrm{scale} $ inter-clump transfer
of rotational momentum can be viewed as a
consequence of the cog-like action of 
clump-surface molecules interacting with one another.
Figure 4 provides a schematic representation of
vortex sheet vorticity on the large BVS scale, the
smaller elemental vortex scale, and on the clump-scale.

\begin{figure}
\centering
\includegraphics[width=6.0in]{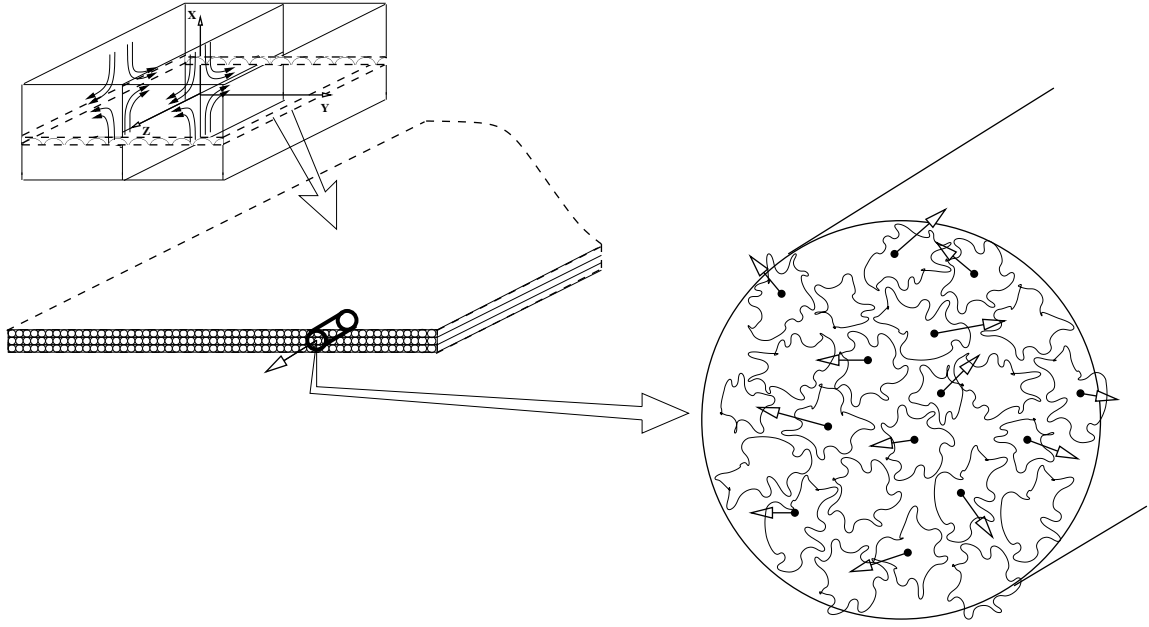}
\caption{On acoustic time scales and on the elemental vortex length
scale, vorticity corresponds to the instantaneous average rotation of $ N_{cl} $
molecules comprising a fluid clump. On these short time and length scales,
and under subsonic conditions, the vorticity field is
three-dimensional, disordered, and strongly diffusive.}
\end{figure} 

Qualitative insight into clump-scale dynamics
can be gained by assuming that a continuum
description applies on scales of order $ \delta_{cl} $ and
$ \tau_{cl} . $
Recognizing that clump-scale gradients in BVS-scale velocities and vorticities
are small
and nondimensionalizing
the full vorticity
transport equation, we obtain:
\begin{equation}\label{vortclumpeqn}
\boldsymbol{\tilde{\omega}}_{\tilde{t}} - \mathrm{Re}^{-1} \tilde{\nabla}^2  
\boldsymbol{\tilde{\omega}} = - \mathrm{Ma} 
\boldsymbol{\hat{u}}
\cdot \boldsymbol{\tilde{\nabla}} \cdot \boldsymbol{\tilde{\omega}} 
\end{equation}
where $ \mathrm{Re}^{-1} = \nu /  ( a_o \delta_{cl} ) $ is a clump-scale Reynolds number,
$ \mathrm{Ma} = u_s/a_o $
is a BVS-scale mach number, $ u_s = O( k x_s) $ is the BVS velocity scale,
$ x_s $ is the BVS-scale sheet position, and where
we have dropped a $ O (\delta_{cl} / x_s) $ term describing stretching of
BVS-scale vorticity by the clump-scale velocity field.

Note first that for
liquids like water, and for clumps having $ O(10^3) $ molecules, $ Re= O(1) . $
Under these circumstances,
and under conditions where
the BVS-scale mach number is small,
(\ref{vortclumpeqn}) 
shows that, relative to diffusion, clump-scale stretching (again, 
dropped from \eqref{vortclumpeqn})
and advection
of vorticity are likewise small.
This is consistent with a detailed analysis
of organization of clump-scale
vorticity by in-sheet hydrodynamic modes below
(section \ref{sec6.3.1});
there, it is shown that diffusional
smoothing via shear and entropy modes
constitute two of three hydrodynamic organizing mechanisms.

Thus, from (\ref{vortclumpeqn}), the following picture of clump-scale
vorticity transport emerges: 
\renewcommand{\labelenumi}{\alph{enumi})}
\begin{enumerate}
\item in low speed 
incompressible flows $ (Ma << 1) , $ 
transport is three dimensional and diffusive, 
\item advection, important on BVS length and time scales,
only emerges on the clump-scale when liquid bulk 
velocities are high, on the order
of the sound speed, and
\item stretching remains weak to the point of nonexistence.
\end{enumerate} 
%%%%%%%%%%%%%%%%%%%%%%%%%%%%%%%%%%%%%%%%%%%%%%%%%%%%%%%%%%%%%%%%%%%%%%%%%%%
\subsubsection{Organization of clump-scale vorticity: hydrodynamic modes}\label{sec6.3.1}
Given that vorticity transport on BVS scales is highly organized and
one dimensional, evolving on these scales under the combined action 
of advection, stretching, and diffusion,
while sub-sheet (clump-scale) transport is 
highly disorganized, three dimensional, and diffusive, it becomes apparent that
some mechanism, 
acting over time and length scales 
long relative to $ \delta_{cl} $ and $ \tau_{cl} , $
organizes disordered clump-scale vorticity. 

We assume that the 
organizing mechanism is associated with
weak hydrodynamic modes superposed
on the bulk, BVS-scale velocity
field. In order to expose the role of
these modes on sub-sheet organization,
we focus on a simpler analog problem,
organization in \textit{unstrained} vortex sheets $(k=0).$

The analysis adapts Mountain's \cite{mountain1966} well known approach
for analyzing the small wave number, low frequency response of 
simple liquids to excitation via
scattered inelastic light and neutron beams \cite{mountain1966, boonyipbook, 
forsterbook}.
In the scattering problem, a probe beam 
interrogates a (nominally) static liquid, with the scattered beam
then detected and analyzed. 

Here, we 
assume that the source
of excited hydrodynamic modes within the vortex sheet derives
from the feature generating sheet-scale vorticity,
e.g., a solid or fluid body moving relative to the initially
static liquid.  
In contrast to the
beam scattering problem
\cite{mountain1966, boonyipbook, forsterbook}, we must
include 
the space- and time-dependent background velocity field, 
$ \mathbf{\bar{u}} = u(y,t) \mathbf{\hat{e}_x}, $
produced by
the vortex sheet.

Model assumptions are as follows: 
\renewcommand{\labelenumi}{\alph{enumi})}
\begin{enumerate}
\item Due to homogeneity of downstream
(\textit{x}-direction) and cross-stream (\textit{z}-direction) 
boundary conditions at the moving, vorticity-generating
boundary or fluid interface,
we assume that hydrodynamic fluctuations
vary only in the cross-sheet (\textit{y-}) direction. 

\item  For the same reason,
we assume that 
only one hydrodynamic shear momentum current, i.e., mass weighted vorticity
component, $ \mu_z (y, t ) = \rho \boldsymbol{\nabla} \mathbf{\times} 
\mathbf{v} \cdot \mathbf{\hat{e}_z} , $ appears and that it is
directed in the cross-stream (\textit{z-}) direction.

\item Consistent with the discussion
above, we assume that hydrodynamic
fluctuations take place on time and length scales
that are short relative to those of the vortex sheet,
and long relative to $ d_o $ and $ \tau_{cl} . $
\end{enumerate}

Thus, express the field variables, $ s , $
$ p , $ $ w , $ and $ \mu_z , $ 
as the superposition of a slowly varying, sheet-scale
component, $ \bar{f} (\bar{y},\bar{t}) , $
and a weak fluctuating sub-sheet component, $ f'(y, t) : $
\begin{equation}\label{feqn}
f(y, \bar{y},t,\bar{t})= \bar{f} (\bar{y}, \bar{t}) + f'(y,t)
\end{equation}
where $ s , $ $ p , $ 
and $ w = \rho \boldsymbol{\nabla} \cdot
\mathbf{v} = \rho \boldsymbol{\nabla} \cdot \mathbf{v'}$ are,
respectively, the entropy, pressure,
and dilatational
momentum currents. In order to expose
sub-sheet-scale processes, sheet-scale time and position
coordinates, $ \bar{y} $ and 
$ \bar{t} , $ are magnified using 
\begin{eqnarray}\label{magcoord}
\bar{y} = \epsilon y \nonumber \\
\bar{t}=\epsilon_1 t \nonumber
\end{eqnarray}
where $ y $ and $ t $ are hydrodynamic-scale 
coordinates and where 
\begin{eqnarray}
\epsilon= d_c / \delta_i \nonumber \\
\epsilon_1 = \tau_{cl} / (d_c/a_o) \nonumber
\end{eqnarray}
Thus, on the sub-sheet, hydrodynamic scale,
time and space derivatives of $ f $ are given by:
\begin{eqnarray}\label{derivf}
\frac{\partial f}{\partial y} = \epsilon \frac{ \partial \bar{f}}{\partial \bar{y}}+
\frac{\partial f'}{\partial y}  \nonumber \\
\frac{\partial f}{\partial t} = \epsilon_1 \frac{ \partial \bar{f}}{\partial \bar{t}}+
\frac{\partial f'}{\partial t}  
\end{eqnarray}
In other words, on the sub-sheet scale, variations 
in sheet-scale fields are small.

Extending the approach in \cite{boonyipbook} to
the problem of hydrodynamic fluctuations within
a planar vortex sheet, we derive the following
system of equations governing
$ s' , $ $ p' , $ $ w' , $ and $ \mu_z' : $
\begin{equation}\label{hydrosystem}
\begin{array}{rcl}
\left[ \frac{\partial}{\partial t} - (k-1)\alpha_T \nabla^2 \right] p'(y,t) +
a_o^2 w'(y,t) 
- \bar{\rho} \beta^{-1} (k-1) \alpha_T \nabla^2 s'(y,t) = & F_1  & = O(\epsilon, \epsilon_1) \\
\left[ \frac{\partial}{\partial t} - (\frac{4}{3} \nu + \nu_B) \nabla^2 \right] 
w'(y,t) 
+ \nabla^2 p'(y,t) = & F_2 & = O(\epsilon, \epsilon_1)  \\
\left[ \frac{\partial}{\partial t} - \alpha_T \nabla^2 \right] s'(y,t) 
- \beta \bar{\rho}^{-1} \alpha_T  \nabla^2 p'(y,t) = & F_3 & = O(\epsilon, \epsilon_1) \\
\left[ \frac{\partial}{\partial t} - \nu \nabla^2 \right] \mu_z'(y,t) = & F_4 & = O(\epsilon, \epsilon_1)   
\end{array}
\end{equation}
where $ k = C_p/C_v $ is the ratio of specific heats, 
$ \alpha_T $ is the thermal diffusivity, 
$ a_o $ is the adiabatic sound speed,
$ \beta $ is the coefficient of thermal expansion, and
$ \nu = \mu/\bar{\rho} $ and $ \nu_B = \mu_B /\bar{\rho} , $ where
$ \mu $ and $ \mu_B $ are the shear and bulk viscosity
coefficients, respectively. 

Prior to listing the $ O(\epsilon, \epsilon_1) $ terms
on the right above, we define
\begin{equation}\label{rhs5}
F_o = \frac{1}{\bar{\rho}\bar{T}} \left[ -(\rho' \bar{T} + T' \bar{\rho} ) 
\frac{\partial \bar{s}}{ \partial t} + \boldsymbol{\Delta'} \cdot 
\boldsymbol{ \pi} + \boldsymbol{\Delta} \cdot \boldsymbol{\pi'} \right]
\end{equation}
where $ \Delta_{ij} = (v_{i,j} + v_{j,i} )/2 $ is the rate of deformation
tensor and $ \pi_{ij} = 2 \eta_1 
\Delta_{ij} + \eta_2 \Delta_{ij} \delta{ij} $ is the 
stress tensor, and $ \eta_1 $ and $ \eta_2 $
are the shear viscosity and dilatational viscosity, respectively
(with $ \eta_1 = \mu $ and $ 2 \eta_1+ \eta_2 = \frac{4}{3} \mu + \mu_B ) . $ 
Primes on $ \Delta_{ij} $ and $ \pi_{ij} $ denote
tensors in the hydrodynamic velocity component; here, 
only derivatives in the cross-sheet $(y-)$ direction appear.
It is seen that products of primed and unprimed versions
of $ \boldsymbol{\Delta} $
and $ \boldsymbol{\pi} $ represent viscous dissipation due to
weak $ O(\epsilon, \epsilon_1) $ interaction between
hydrodynamic-scale and sheet-scale velocity fields. 

Given $ F_o , $ the $ O(\epsilon, \epsilon_1) $ 
terms above can be expressed as: 
$ F_1 = (k-1) \bar{\rho} \beta^{-1} F_o , $
$ F_2 = - \mathbf{\bar{v}}_t  \cdot \rho' - \bar{\rho} v_{j,i'} \bar{v}_{i,j} , $
$ F_3 = F_o , $ and $ F_4 = - \bar{\rho}^{-1} \boldsymbol{\nabla} \mathbf{\times}
( \rho' \mathbf{\bar{v}}_t ) - \boldsymbol{\nabla} \mathbf{\times} [ \mathbf{v'}
\cdot \boldsymbol{\nabla} \mathbf{\bar{v}} ] . $

Importantly, the $ O(1) $ left hand side of
(\ref{hydrosystem}) is, with the exception of the appearance of one
rather than two shear momentum currents, identical to that
describing the hydrodynamic response of a nominally static liquid
to beam scattering. Thus, focusing on the leading order 
problem, we can immediately apply the results
of the scattering problem to sub-sheet organization in unstrained
vortex sheets.
Taking the Laplace-Fourier transform of the above system
\begin{displaymath}
\hat{f'}_k = \int_0^{\infty} \int_{-\infty}^{\infty}
e^{i \mathbf{k} \cdot \mathbf{x} } e^{-st} f'(\mathbf{x}, t) d \mathbf{x} dt
\end{displaymath}
with $ \mathbf{x}= y \mathbf{\hat{e}_y} , $
and enforcing the solvability condition 
leads to
\begin{equation}\label{ftsystem}
\left| \begin{array}{cccc}
[s + (k-1) \alpha_T k^2] & a_o^2 & \bar{\rho} \beta^{-1} (k-1) \alpha_T k^2 
& 0 \\
- k^2 & [s + \nu_l k^2 ] & 0 & 0 \\
\bar{\rho}^{-1} \beta k^2 & 0 & [s + \alpha_T k^2] & 0 \\
0 & 0 & 0 & [s + \nu k^2] 
\end{array} \right| = 0
\end{equation}

From the original system (or (\ref{ftsystem})), it is clear that
the cross-sheet shear momentum current is uncoupled from the 
other three modes; from (\ref{ftsystem}), the
associated dispersion relation
\begin{equation}\label{hydmode3}
s_4 = - \nu k^2 
\end{equation}
shows that this
mode is purely diffusive.
The remaining modes follow from (\ref{ftsystem})
and are given by \cite{mountain1966, boonyipbook}:
\begin{eqnarray}
s_1 & = - \alpha_T k^2 \label{hydmode1} \\
s_{2,3} & = \pm i a_o k - A_o k^2 \label{hydmode2}
\end{eqnarray}
where $ A_o = [ \nu_l + (k-1) \alpha_T]/2 , $
with $ \nu_l = \frac{4}{3} \nu + \nu_B . $

Thus, from (\ref{hydmode1}) and (\ref{hydmode2}), we
surmise that, beyond
cross-sheet diffusion of clump-scale vorticity,
two additional sub-sheet organizing mechanisms act (respectively):
thermal smoothing
via a purely diffusive cross-sheet entropy mode, and
acoustic smoothing via weakly damped, oppositely-directed, cross-sheet 
acoustic modes (with the rate of damping determined
by $ A_o ). $ 

Physically, in the cross-sheet
direction, acoustic dilatational waves
tend to push out-of-plane vorticity fluctuations into a planarized
configuration; likewise, within the in-sheet 
plane, wave reflection between neighboring vorticity fluctuations tend to align
the fluctuations in the $ z-\mathrm{direction} . $
Superposed on these acoustic mechanisms, diffusional smoothing of 
clump-scale vorticity as well diffusion of friction-generated
thermal energy provide additional, strongly organizing effects.
The mechanisms
are depicted in figure 5; see the caption for further discussion. 

\begin{figure}
\centering
\includegraphics[width=5.0in]{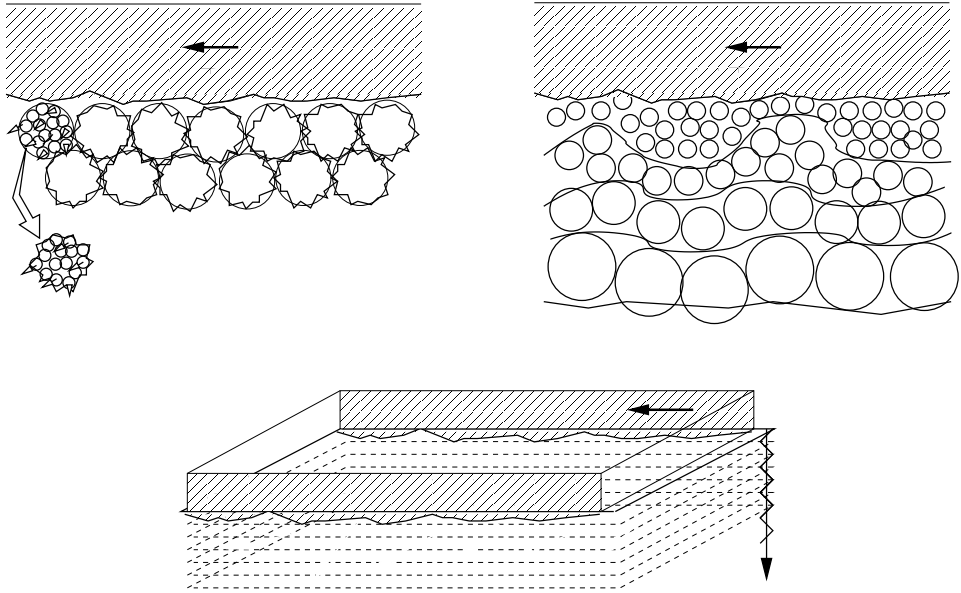} 
\caption{Organization of disordered clump-scale
vorticity is mediated by three hydrodynamic modes:
a diffusive, cross-sheet entropy (thermal) mode (upper left),
a diffusive, cross-sheet shear (vorticity) mode (upper right), and
a pair of counter-propagating, weakly damped, cross-sheet
acoustic modes (bottom). We depict the thermal mode in terms of
the thermal degradation (dispersion) of in-sheet clumps. The
vorticity mode can likewise be viewed as the hydrodynamic-time-scale
clump-to-clump transfer of rotational inertia. On these time scales,
clumps experience significant dispersion so that rotational
transport takes place by both the cog-like interaction
of adjacent 'thermal bumps', and gross intermixing
of neighboring clumps.}
\end{figure} 

%%%%%%%%%%%%%%%%%%%%%%%%%%%%%%%%%%%%%%%%%%%%%%%%%%%%
\subsubsection{Elemental vortex sheets defined}\label{sec6.3.2}
Having developed a physical picture
of sub-BVS vorticity, we 
can identify a reasonable
physical embodiment of the sub-BVS scale stochastic processes 
defined by equation (\ref{vortexchi}):
elemental vortex sheets are defined as 
planar vorticity layers of thickness $ O(\delta_{cl} ) , $
which evolve stochastically within any given BVS, on the long BVS time-scale. 

Two features suggest that EVS's can be taken as real, 
rather than quasi-physical,
entities. 
First, since hydrodynamic acoustic, shear, and entropy
modes organize, on long BVS length and time scales,
disordered small- (clump-) scale vorticity,
it is clear that these long-acting modes are likewise capable
of organizing 
$ O(\delta_c) $ layers of vorticity.
Thus, on BVS time scales,
it is physically reasonable to view, \textit{at any instant,} 
a Burger's vortex sheet, or any vortex sheet, as
being composed of $ O(\delta_i/\delta_{cl}) $ elemental vortex sheets.
Additionally, 
it is likewise reasonable to expect
that due to long-time-scale organization of short-time-scale vorticity,
$ O(\delta_{cl}) $ layers of vorticity evolve stochastically,
\textit{en mass}, over long BVS time increments, $ \Delta t . $    
%%%%%%%%%%%%%%%%%%%%%%%%%%%%%%%%%%%%%%%%%%%%%%%%%%%%%%%%%%%%%
\subsection{Vortex sheet evolution in non-random strain rate fields}\label{sec6.4.1}
Prior to discussing vortex sheets in random strain rate fields, we briefly 
consider physically interesting aspects of the solution which apply
to non-random strain fields. 
Thus, using (\ref{finalvortexsoln}) in (\ref{vortexgenlsoln}), one can 
write down the vorticity field produced by $ N_x $ initially discrete 
Burger's vortex sheets: 
\begin{equation}\label{nvortexsheets}
\Omega (x,t) = \sum_{i=1}^{N_x} \frac{\Delta v_i}{\sqrt{4 \pi \nu p(t)}}
\exp \big[\frac{ -(x-x_i e^{-h(t)})^2}{4 \pi \nu p(t)} \big] ,
\end{equation}
where again, $ \Delta v_i $ is the incremental change in the streamwise 
velocity at $ x_i . $

Several observations follow from this solution. 
First, from (\ref{vortexmeanpos}) and (\ref{vortexsigma}),
the mean position and diffusive spread of the $ i^{th} $ sheet,
viewed as the stochastic evolution
of a cloud of elemental vortex sheets,
are given respectively by:
\begin{equation}\label{sheetimeanposition}
< \chi_i (t) >_x =  x_i \exp \big({-h(t)} \big)
\end{equation}
and
\begin{equation}\label{sheetisigma}
< (\chi_i (t) - <\chi_i(t)>_x)^2 >_x = 2 \nu p(t)
\end{equation}
where $ h(t) $ and $ p(t) $ are given by (\ref{heqn}) and (\ref{pdefn}).
Thus, in non-random strain rate fields (and due to the linearity of the problem),
each sheet (in the collection of $ N_x $ sheets) migrates
toward $ x = 0 , $ each diffusively spreading at a spatially uniform rate;
at large times, all $ N_x $ sheets coalesce at $ x = 0 , $
with the thickness of the collection, $ \sqrt{2 \nu p(t)} , $ identical to that
of a single sheet.
In the case where $ k(t) $ is constant, the above mean and variance
simplify to
those for the equivalent OU process considered in \cite{gardiner}.
Likewise, 
identifying 
$ \sqrt{< (\chi_i (t) - <\chi_i(t)>_x)^2 >_x} $ as the thickness of sheet $ i , $
we obtain the constant strain rate expressions for sheet
position and thickness given, e.g., in \cite{sherman, saffman}.

Second, and considering the long-time strength of the composite sheet
in the case where $ k $ is either (a positive) constant or approaches 
a long-time limit, $ k \rightarrow k_{\infty} , $
$ h(t) \rightarrow \infty $ and $ p(t) \rightarrow 1/(2 k_{\infty} ) , $
and the strength of the coalesced collection of $ N_x $
sheets, from (\ref{nvortexsheets}), is the sum of the initial strengths:
\begin{equation}\label{finalstrength}
v_{\infty} = \sum_{i=1}^{N_x} \Delta v_i 
\end{equation}
Thus, for example, when the strength of one 
of the initial sheets dominates all others, corresponding to
a locally large change in the initial stream-wise velocity profile,
(\ref{finalstrength}) shows that 
the final sheet strength is largely determined by the strength of this sheet.
In contrast, in situations where the initial streamwise 
velocity profile takes on positive and negative magnitudes,
and more specifically, is characterized by a limited number of
Fourier modes, (\ref{finalstrength}) shows that due to vorticity cancellation
the final strength is small.
%%%%%%%%%%%%%%%%%%%%%%%%%%%%%%%%%%%%%%%%%%%%%%%%%%%%%
\subsection{Single and continuous sets of Burger's vortex sheets in random strain 
rate fields}\label{sec6.5}
Focusing first on the dynamics of a single vortex sheet
$ (N_x =1 $ in (\ref{nvortexsheets})), we note that since equations 
(\ref{nvortexsheets}), (\ref{sheetimeanposition}),
and (\ref{sheetisigma}) apply to individual realizations of
$ k(t) , $ then the corresponding ensemble averaged instantaneous vorticity,
sheet position, and sheet thickness, taken over
$ k , $ can be expressed respectively as:
\begin{equation}\label{ranvortexsheet}
\big< \Omega (x,t) \big>  = \big< \frac{\Delta v}{\sqrt{4 \pi \nu p(t)}}
\exp \big[\frac{ -(x-x_o e^{-h(t)})^2}{4 \pi \nu p(t)} \big] \big>
\end{equation}
\begin{equation}\label{ransheetmeanposition}
\big< < \chi (t) >_x \big> =  x_o \big< \exp \big({-h(t)} \big) \big>
\end{equation}
and
\begin{equation}\label{ransheetsigma}
\big< < (\chi (t) - <\chi (t)>_x)^2 >_x \big> = 2 \nu \big< p(t) \big>
\end{equation}
where $ h(t) $ and $ p(t) $ are again given by (\ref{heqn}) and (\ref{pdefn}), 
and $ \Delta v $ is the incremental (step-function) change in the streamwise 
velocity at $ x_o . $ 

Explicit formulas for these quantities can be obtained
when the strain rate field is statistically stationary,
has positive mean, and a gaussian random component:
\begin{equation}\label{krecipe}
k(t) = k_o + k'(t)
\end{equation}
where, for stability, we assume a condition like $ k_o^2 > \alpha \big< (k')^2 \big> , $
where $ \alpha $ is an undetermined positive number. 
Under these conditions 
\cite{gelfand1964}:
\begin{equation}\label{gaussianstatid1}
\Big< \exp \big[ - \int_0^t k'(s) ds \big] \Big> =
\exp \big[ \frac{1}{2} \int_0^t \int_0^t \big< k'(s) k'(s') \big> ds ds' \big] 
\end{equation}

Inserting the correlation function, $ R , $ 
and using the property
\begin{equation}\label{kcorrfn}
R(|\tau|) = \big< k'(t'+\tau) k' (t') \big>
\end{equation}
appropriate to stationary processes, allows restatement of
(\ref{gaussianstatid1}) in the form:
\begin{equation}\label{gaussianstatid2}
\Big< \exp \big[ - \int_0^t k'(s) ds \big] \Big> =
\exp \big[ \frac{1}{2} \int_0^t \int_0^t R ( |s - s'|) ds ds' \big] 
\end{equation}
In turn, the identity \cite{ave1994}
\begin{equation}\label{rid}
\frac{1}{2} \int_0^t \int_0^t R(|s -s'|) ds ds' = \int_0^t (t-s) R(|s|) ds
\end{equation}
is useful when evaluating the right side of (\ref{gaussianstatid2}).

Thus, inserting (\ref{gaussianstatid2}) and (\ref{rid}) in (\ref{ransheetmeanposition}) 
yields the following formula for the (single)
sheet's ensemble average position,
applicable to random strain fields
of the form in (\ref{krecipe}):
\begin{equation}\label{ransheetmeanpositiongauss}
\big< < \chi (t) >_x \big> =  x_o e^{-k_o t} \exp \big[ \int_0^t (t-s) R(|s|) ds \big]
\end{equation}

In order to obtain the corresponding single sheet spread,
we use (\ref{pdefn}) to write 
\begin{equation}\label{pgauss}
\big< p(t) \big> = e^{-2 k_o t} \big< \exp{ \big[ -2 \int_0^t k'(t') dt' \big] } \cdot
\int_0^t \Big( e^{2 k_o t'} \exp{ \big[ 2 \int_0^{t'} k'(s') ds' \big] } \Big) dt' \big>
\end{equation}
and assume that the two terms separated by the center dot
represent independent random variables:
\begin{equation}\label{pgauss1}
\big< p(t) \big> = e^{-2 k_o t} \big< \exp{ \big[ -2 \int_0^t k'(t') dt' \big] } \big>
\cdot \big< \int_0^t \Big( e^{2 k_o t'} \exp{ \big[ 2 \int_0^{t'} k'(s') ds' \big] } \Big) dt' \big>
\end{equation}
or
\begin{equation}\label{pgauss2}
\big< p(t) \big> = e^{-2 k_o t} \big< \exp{ \big[ -2 \int_0^t k'(t') dt' \big] } \big>
\cdot \int_0^t \Big( e^{2 k_o t'} \big< \exp{ \big[ 2 \int_0^{t'} k'(s') ds' \big] } \big>
\Big) dt' 
\end{equation}
Using (\ref{gaussianstatid2}) and (\ref{rid}) to evaluate the
expectations in (\ref{pgauss2}) and inserting the results
in (\ref{ransheetsigma}) then yields:
\begin{equation}\label{ransheetspreadk}
\big< < (\chi (t) - <\chi (t)>_x)^2 >_x \big> 
= 2 \nu e^{-2 k_o t} 
\exp \big[ 4 \int_0^t (t-s) R(|s|) ds \big] 
\cdot \int_0^t \Big( e^{2 k_o t'} \exp \big[ 4 \int_0^{t'} (t'-s')R(|s'|) ds' \big] \Big)
dt'
\end{equation}

Considering, for illustration, the case where the random strain field is
delta correlated in time 
\begin{equation}\label{kcorrelation}
\big< k'(t'+\tau) k'(t') \big> = \tilde{k} \delta (\tau)
\end{equation}
where $ \tilde{k} $ is a positive constant (determined, e.g., by experiment),
one finds that sheet spread no longer approaches a fixed long-time 
limit (as in the case of deterministic strain rate), but rather 
exhibits exponential, i.e., superdiffusive
\cite{majdakramer} growth:
\begin{equation}\label{ransheetspreaddelta}
\big< < (\chi (t) - <\chi (t)>_x)^2 >_x \big> = \frac{ \nu}{k_o + 2 \tilde{k}}
\exp \Big(8 \tilde{k} t \big[ 1 - \exp \big( -( 2 k_o + 4 \tilde{k}) t \big) \big] \Big)
\end{equation}
In the limit of weak
random strain, specifically $ \tilde{k} t << 1 $
and $ k_o >> \tilde{k} , $ (\ref{ransheetspreaddelta}) simplifies to
the constant strain (non-random) solution \cite{gardiner}:
\begin{equation}\label{limitsmallkprime}
\big< < (\chi (t) - <\chi (t)>_x)^2 >_x \big> = \frac{ \nu}{k_o}
\big[ 1 - \exp \big( - 2 k_o t \big) \big]
\end{equation}
[See section \ref{sec6.6.1} below for a brief description of
the conditions necessary for introducing an assumption of
delta correlated strain rates.]
\subsubsection{Continuous collections of Burger's vortex sheets: 
Feynman-Kac solution}\label{sec6.5.1}
Turning to the case of multiple Burger's vortex sheets, we
consider the general case where the initial streamwise velocity
component exhibits a continuous variation in the $ x-$ direction,
\begin{equation}\label{vintial}
v(x,0) = v_o(x) 
\end{equation}
Here, a representative solution for the mean continuous vorticity,
$ \big< \omega(x , t) \big> , $ can be obtained using the Feynman-Kac
solution \cite{feynmankatzref} of the backward-time version of the vorticity equation 
(\ref{burgervortexgoveqn}):
\begin{equation}\label{fkvortexsoln}
\big< \Omega(x,t) \big> =
\big< E_{\chi(t)=x,t} \big[ \Omega (\chi (0),0) \big] \big> \cdot
\big< e^{h(t)} \big>
\end{equation}
where $ E_{\chi(t), t} $ denotes the expectation associated with
the stochastic process $ \chi $ sampling the initial condition,
$ \Omega(x,0) . $

As a quick check of this formula, 
we note that 
when $ k(t) = k_o $ is non-random and constant and $ v_o = c_o x , $ i.e.,
the initial streamwise velocity is of Couette form, 
the solution in (\ref{fkvortexsoln}) yields 
\begin{equation}\label{constinitvo}
\Omega (x,t) = c_o e^{k_o  t } 
\end{equation}
This solution in turn satisfies the vorticity transport equation, 
(\ref{burgervortexgoveqn}), along with the initial condition, $ \Omega(x,0) = c_o . $
Equation (\ref{constinitvo}) holds under conditions
where the time scale for setting up the Couette profile, say $ \tau , $
is short relative to the  strain field time scale, $ \tau_s = k_o^{-1} . $
On the longer scale, diffusion of vorticity
saturates out quickly and the appropriate initial condition,
to order $ k_o^{-1} / \tau , $ is $ \Omega (x , 0 ) = c_o . $
Under these conditions, the initial vorticity, $ c_o , $ 
is exponentially amplified by stretching.

As shown immediately below, in the case where $ k(t) $ is random and $ v_o $ again equals $ c_o x , $
the Feynman-Kac solution in (\ref{fkvortexsoln}) 
coincides with the leading order low-viscosity (high Reynolds number) 
forward-time solution of (\ref{burgervortexgoveqn}). The paradox of the viscous solution
coinciding with the inviscid solution is resolved by
requiring (in analogy with the deterministic strain case immediately above),
that for every realization of $ k(t) , $ 
the initial (linear) streamwise velocity profile forms
on time scales much shorter than say, $ | k^{-1} (t) |_{max} ; $
for each realization, diffusive effects again saturate out over the short time scale.
We now briefly consider the latter, low viscosity problem. 
%%%%%%%%%%%%%%%%%%%%%%%%%%%%%%%%%%%%%%%%%%%%%%%%%%%%%%%%%%%%
\subsection{Single-sheet and collective behavior under inviscid 
conditions}\label{lownulimit}\label{sec6.6}
In the case of single sheets, significant questions include the effects of random strain
on time-varying mean (ensemble averaged) sheet strength, position, and spread,
while in the case of sheet collections, the key question centers on
evolution of the ensemble average (position and time-dependent) vorticity.
Attention is again focused on the case where $ k(t) $ is given by
(\ref{krecipe}) with $ k'(t) $ again a stationary zero-mean gaussian 
process.
  
Considering first single sheets, the position of the sheet 
under the action of an individual realization of $ k (t ) ,$
\begin{equation}\label{chisingle}
\chi (t) = e^{-h(t)} \chi(0)
\end{equation}
follows either from the SDE governing the
motion of elemental vortex sheets, (\ref{vortexchi}), with
the stochastic term suppressed, or by noting from the 
MOC solution for $ \Omega (x, t) $ (using the inviscid form of
(\ref{burgervortexgoveqn})) that any given sheet evolves along characteristics
given by (\ref{character1}). In the case of elemental sheets,
since no diffusion takes place, all EVS's within the Burger's sheet
evolve identically under the action of any given realization of $ k(t) . $
In the following, we express the sheet's initial
position as $ \chi(0) = x_o . $ 

Given $ \chi(t) , $ the ensemble average sheet position over
the random strain rate field can be calculated using 
(\ref{gaussianstatid2}) and (\ref{rid}): 
\begin{equation}\label{meansheetpos}
\big< \chi(t) \big> = x_o e^{-k_o t} \big< \exp \big[ - \int_0^t k'(s) ds \big] \big> =
x_o e^{-k_o t} \exp \left[\int_0^t (t-s) R(|s|) ds \right]
\end{equation}
Since $ (t-s) R(|s|) $ is positive for $ s < t , $
then it is seen that any stationary random strain rate field
inhibits sheet migration toward the origin, $ x = 0 . $

Given $ \big< \chi(t) \big> , $ the sheet spread (or squared thickness), 
$ \sigma_{\chi}^2 , $ can 
again be calculated:
\begin{equation}\label{meansheetspread}
\sigma_{\chi}^2 = \Big< (\chi(t) - \big< \chi \big>)^2 \Big> =
x_o^2 e^{-2 k_o t} \big[ \big< \exp -2 \int_0^t  k'(t') dt' \big> 
 - \big< \exp - \int_0^t k'(t') dt' \big>^2 \big]
\end{equation}
or
\begin{equation}\label{meansheetspread1}
\sigma_{\chi}^2 = 
x_o^2 e^{-2 k_o t} \big[ \exp[ 4 \int_0^t R(|s|) (t-s) ds ]  
 - \exp[ 2 \int_0^t R(|s|)(t-s) ds ] \big]
\end{equation}
%%%%%%%%%%%%%%%%%%%%%%%%%%%%%%%%%%%%%%%%%%%%%%%%%%%%%%%%%%%%%%5
\subsubsection{Inviscid vortex evolution in a rapidly decorrellating
random strain
rate field}\label{sec6.6.1}
This example briefly considers the spread of a single Burger's vortex sheet 
under conditions where
the correlation time, $ \tau_c = \int_0^{\infty} \left< k'(t) k'(0) \right> dt /
\mathrm{var} (k') , $ of the random strain field, $ k'(t) , $ is
short relative to the time scale, $ \tau_b , $ associated with the vorticity-inducing
flow, $ v(x,t) . $ Thus, define a normalized correlation function
$ \eta(\tau) = \left< k'(\tau) k'(0) \right> /c_1^2 , $
where $ c_1^2 $ is a normalization factor obtained from $ \int_{-\infty}^{\infty}
\eta (\tau) d \tau = 1 . $ In addition, let $ \epsilon = \tau_c / \tau_b << 1 , $
and renormalize the correlation function to the  longer BVS time scale: 
$ \eta_{\epsilon} (\tilde{\tau} ) = \epsilon^{-1} \eta(\epsilon^{-1} \tau) , $
where $ \tilde{\tau} = \tau / \epsilon . $ Since $ \eta_{\epsilon} (\tilde{\tau} )
\rightarrow \delta (\tilde{\tau} ) $ as $ \epsilon \rightarrow 0 $
(and $ \tilde{\tau} = O (\tau_b ) ), $
then it is clear that the assumption of delta correlated statistics on the BVS
time scale becomes increasingly valid as the ratio of the 
$ \tau_c / \tau_b $ becomes increasingly small.

Thus, let $ k'(t) $ be delta correlated. In this case,
(\ref{meansheetpos}) yields:
\begin{equation}\label{meansheetpos1}
\big< \chi(t) \big> = x_o e^{(-k_o + \tilde{k}) t}
\end{equation}
Although this suggests that for large enough random strain, $ \tilde{k} > k_o , $
the mean position of the BVS can track away from the origin, $ x=0 , $
this cannot be confirmed without detailed consideration of sheet
stability under random strain rates.

Under the same assumption, a variety of interesting long-time
behaviors emerge, depending on the size of $ \tilde{k} $ relative
to $ k_o : $
\begin{equation}\label{meansheetspread2}
\sigma_{\chi}^2 = 
x_o^2 e^{-2 k_o t} \big[ e^{4 \tilde{k}t} - e^{2 \tilde{k}t} \big]
\end{equation}
In particular: 
\renewcommand{\labelenumi}{\roman{enumi})}
\begin{enumerate}
\item when $ \tilde{k} < k_o/2 , $ corresponding to mean strain dominating the random component,
sheet thickness goes to $ 0 $
(i.e., $ \sigma_{\chi}^2 \rightarrow 0) $ as
$ t \rightarrow \infty ; $

\item when $ \tilde{k} > k_o/2 , $ sheet thickness grows without bound 
$ ( \sigma_{\chi}^2 \rightarrow \infty) $ as
$ t \rightarrow \infty ; $ 

\item in the special case where $ \tilde{k} = k_o/2 , $ a fixed, non-zero thickness 
can be achieved,
$ \sigma_{\chi} \rightarrow x_o $
as $ t \rightarrow \infty , $ indicating, both here and in the other cases,
that the random strain functions as an effective agent for diffusion. 
\end{enumerate}
\noindent In each case, the increase in sheet spread amplitude with initial position, $ x_o^2 , $
reflects the increase in random velocity amplitude, $ v'(x,t) = k'(t) x' , $
with $ x . $ Finally, and once again, cases ii) and iii) are subject to the proviso
concerning sheet stability in increasingly random strain fields.
%%%%%%%%%%%%%%%%%%%%%%%%%%%%%%%%%%%%%%%%%%%%%%%%%%%%%%%%%%%%%%%%
\subsubsection{Single sheet and continuous 
vorticity in the inviscid limit}\label{sec6.6.2}
Considering the averaged evolution of the vorticity, $ \big< \Omega (x,t) \big> , $
we first introduce the limit $ \nu \rightarrow 0 $ into the single sheet solution 
(\ref{finalvortexsoln}) (which is applicable to
individual realizations of $ k (t) ), $
to obtain
\begin{equation}\label{smallnuvortex}
\Omega (x, t) = \Delta v \delta (x - x_o e^{-h(t)} )
\end{equation}
where again $ h(t) $ is in defined in (\ref{heqn}).
Physically, a vortex sheet initiated at $ x_o $
simply advects with the random strain field.
Expressing the delta function in its Fourier representation form,
followed by evaluation of the ensemble average
then yields:
\begin{equation}\label{inviscidvortex}
\big< \Omega (x,t) \big> = \frac{\Delta v}{2 \pi} \int_{-\infty}^{\infty}
e^{imx} \big< \exp[ -imx_o e^{-k_o t} e^{-\int_0^t k'(t') dt'} \big> dm
\end{equation}

We refrain from analyzing this expression in depth, but
note in the case where mean strain dominates
random strain, $ k_o >> \tilde{k} , $ the long-time limit of (\ref{inviscidvortex}) yields
\begin{equation}\label{inviscidvortex1}
\big< \Omega (x,t) \big> = \frac{\Delta v}{2 \pi} \int_{-\infty}^{\infty}
e^{imx} dm = \Delta v \delta (x) 
\end{equation}
Thus, when strain rate is strongly deterministic, and under inviscid conditions,
the initial, infinitely thin vortex sheet remains infinitely thin, migrates to
$ x = 0 , $ and becomes infinitely strong, all physically reasonable results.

Turning next to the inviscid evolution of a continuous set of vortex sheets,
and focusing initially on a single realization of random
$ k(t) , $ we first 
write the discrete multi-sheet solution corresponding to (\ref{nvortexsheets}) as 
\begin{equation}\label{nvortexsheets1}
\Omega (x,t) = \sum_{i=1}^{N_x} \Delta v_i \delta(x - x_i e^{-h(t)})
\end{equation}
and then re-express this as an integral:
\begin{equation}\label{nvortexsheets2}
\Omega (x,t) = \int_{-\infty}^{\infty} \frac{\partial V(x', t=0)}{\partial x'}
\delta (x - x' e^{-h(t)}) dx'
\end{equation}
Integrating then gives
\begin{equation}\label{nvortexsheets3}
\Omega (x,t) = e^{h(t)} \frac{\partial}{\partial x} v (x_{init}(t),0)
\end{equation}
where 
\begin{equation}\label{xinit}
x_{init}(t) = x e^{h(t)} 
\end{equation}
is the initial $ x $ position (at $ t = 0 ) $
of the characteristic passing through $ x $ at time $ t ; $ refer to equation (\ref{chisingle}).
Thus, for any given realization of $ k , $
the initial vorticity, generated at $ x_{init} (t) $ and having magnitude 
\begin{displaymath}
\frac{\partial}{\partial x} v(x_{init}(t),0)
\end{displaymath}
is amplified (via stretching) by a factor of $ e^{h(t)} . $

The ensemble average vorticity assumes the form
\begin{equation}\label{nvortexsheets4}
\big< \Omega (x,t) \big>  = \big< e^{h(t)} \frac{\partial}{\partial x'}v(x_{init}(t),0) \big>
\end{equation}
The consistency of this inviscid solution is shown by noting
that the representative \textit{general} (viscous) solution obtained via
the Feynman-Kac approach, equation (\ref{fkvortexsoln}),
simplifies to (\ref{nvortexsheets4}) when the stochastic (i.e., viscous) term
in the SDE governing $ \chi(t) $ is set to zero. In other words, for any realization of $ k(t) , $
the stochastic processes
used to construct the FK solution track along the inviscid characteristics, $ dx = -k(t) x dt , $
so that 
\begin{displaymath}
E_{\chi(t)=x,t} \Omega (\chi(0), 0 ) = \Omega( x_{init}(t),0) = 
\frac{\partial}{\partial x} v(x_{init}(t),0)
\end{displaymath}
yielding (\ref{nvortexsheets4}).

As a simple example, in the case
where the initial streamwise velocity is of Couette form, $ v(x', t=0) = \Omega_o x' , $
where $ \Omega_o (= c_o $ above) is constant:
\begin{equation}\label{nvortexsheets5}
\big< \Omega (x,t) \big>  = \Omega_o \big< e^{h(t)} \big> 
\end{equation}
When $ k(t) $ is
given by (\ref{krecipe}) with, e.g., a delta correlated
random component $ k'(t) , $ 
(\ref{nvortexsheets5}) assumes the form:
\begin{equation}\label{collectsoln}
\big< \Omega (x,t) \big>  = \Omega_o e^{(k_o + \tilde{k})t}
\end{equation}
Hence, in this simple example, under inviscid (i.e., high Reynolds number) conditions,
\textit{continuous,} initially constant vorticity is amplified
not only by the mean strain field, $ k_o , $ but also by
the random component. 
%%%%%%%%%%%%%%%%%%%%%%%%%%%%%%%%%%%%%%%%%%%%%%%%%%%
\section{Concluding remarks} 
Three themes are pursued. First, 
by comparing the representative stochastic solution
of a linear, nonhomogeneous, drift-diffusion 
problem 
against the corresponding Green's function solution,
we obtain a set of equalities relating
stochastic expectations in the former to
Green's function convolutions in the latter.
Importantly, 
these equalities expose a framework
within which
stochastic 
and Green's function methods can be applied in concert
to a range of problems.  
In broad terms, and as illustrated above,
the framework allows exploitation of two 
generic constructs -- delta functions
and Wiener processes -- for mathematically and physically
modeling and probing an array of linear and nonlinear,
deterministic and random problems.

The second theme centers on testing application of 
time-incremental GF's to solution of nonlinear
evolution problems.
Two canonical problems,
Burger's equation
and the nonlinear Schr\"odinger equation, provide test beds.
The first Test Case, focused on the 
simplest embodiment of
nonlinear drift-diffusion problems, leads to the following
observations:
\renewcommand{\labelenumi}{\roman{enumi})}
\begin{enumerate}
\item Transforming from an incremental to nonincremental
solution rests on elimination of nonlinearity
from the incremental GF.
Here, the transformation is accomplished via the following procedure: 

\renewcommand{\labelenumi}{\alph{enumi})}
\begin{enumerate}
\item guess a transform, $ \phi(u) , $
parameterized in the unknown $ u , $
\item force the evolution of $ \phi(u) $ to be purely diffusive, and
\item attempt to match the evolution of $ u $ that emerges
from the diffusion of $ \phi(u) , $ equation \eqref{cole1}, to the original
nonlinear evolution equation governing $ u, $ equation \eqref{burgers}. 
\end{enumerate}
\renewcommand{\labelenumi}{\roman{enumi})}
\item Proper choice of the time step, embodied, for example,
in equations \eqref{timestepcond1} and \eqref{timestepcond2},
allows approximate
solution of the otherwise difficult-to-solve
backward adjoint problem. This step is essential
to obtaining an analytical incremental GF.
\item The procedure outlined in i) and ii) allows \textit{derivation} of
the Cole-Hopf ansatz. Moreover, we
anticipate that a similar approach can applied to
other nonlinear evolution problems.
\end{enumerate}
The second Test Case, an example of
nonlinear parabolic wave propagation,
illustrates application of asymptotic approaches for
transforming from 
incremental to non-incremental solutions;
here, a 
well-known soliton solution is recovered.

The last theme revolves around
use of the GFSM as a tool for probing 
physical features in problems characterized by some element of randomness.
Here, a thorough
investigation of single, multiple, and continuous sets of
Burger's vortex sheets evolving in deterministic and random
strain rate fields, under viscous and inviscid flow conditions, is presented.
The main results are as follows: 
\renewcommand{\labelenumi}{\roman{enumi})}
\begin{enumerate}
\item For delta function initial conditions, the vorticity transport equation
assumes the form of a Fokker-Planck equation of an Ornstein-Uhlenbeck
process. 

\item Correspondingly, the evolution of a Burger's vortex sheet, i.e., the
movement of its mean position and its time-varying spread,     
can be viewed as the
evolution of a constituent cloud of \textit{elemental vortex sheets,}
governed by an OU stochastic differential equation.

\item A physical picture of EVS's is developed by first focusing
on the short time-scale vorticity evolution of particle clumps.
On short acoustic time scales, sub-sheet vorticity is three-dimensional, disordered,
and strongly
diffusive. 

\item Considering the fundamental question of how disordered, clump-scale
vorticity becomes organized on long BVS time scales,
we study an analog problem of organization within
non-strained planar vortex sheets. In this case, 
an analysis of sub-sheet hydrodynamic modes
suggests that organization is mediated by a combination
of damped, cross-sheet acoustic modes, a diffusional
cross-sheet shear mode, and a diffusional cross-sheet 
entropy mode. 

\item A number of analytical results 
describing the motion and spread of individual, discrete collections,
and continuous sets of Burger's vortex sheets, evolving
within deterministic and random strain rate fields,
under both viscous and inviscid conditions, are also obtained.
\end{enumerate}

%%%%%%%%%%%%%%%%%%%%%%%%%%%%%%%%%
%%%%%%%%%%%%%%%%%%%%%%%%%%%%%%%
\renewcommand{\theequation}{A-\arabic{equation}}
\setcounter{equation}{0}
\section*{Appendix: Definitions; notes regarding incremental GF's}
Definitions of forward and backward time
coordinates and forward and backward evolution and adjoint equations
are first given.
Four important 
points regarding determination of time-incremental
Green's functions are then highlighted. 

\vspace{1.2cm}

\noindent \textbf{Forward and backward time coordinates and equations}\\
Coordinates in the forward, i.e., physical time 
direction are denoted as $ t $ or $ t' ; $
the backward time coordinate is denoted as $ s $ or $ s' , $
A drift-diffusion
(or drift-diffusion-like) evolution problem is
of \textit{forward time form} if: i) the 
signs on the time derivative and diffusion terms in equation
(\ref{nonlinearpattern})
differ, and ii) the evolution is initiated from 
some known initial state. Likewise,
an evolution problem is of \textit{backward time form}
if: i) signs on the time derivative and diffusion terms in
(\ref{nonlinearpattern})
are the same, and ii) the evolution \textit{ends} at
some known final state. 

\vspace{1.2cm}

\noindent \textbf{Notes on incremental GF's}\\
We highlight four points.

First, the incremental Green's function, $ G, $ must
meet the following
requirements:\\
i) It must satisfy the adjoint problem
associated with the
evolution problem of interest. \\
ii) For \textit{forward form} adjoint problems, again associated with
backward time evolution problems (both stated in terms of the backward time coordinate, $ s' ) , $
$ G $ must behave as $ \delta(\mathbf{x'} - \mathbf{x}) $
as $ s' \in [s_j, s_{j+1}) $
approaches $ s_{j} ; $ here, $ \mathbf{x'} $ and $ \mathbf{x} $
are variable and fixed points. Likewise, for \textit{backward form}
adjoint problems, associated with forward time evolution problems,
both stated in terms of the forward time coordinate, $ t' , $
$ G \rightarrow \delta (\mathbf{x'} - \mathbf{x}) $
as $ t' \rightarrow t_{j} $ (where
$ t_{j} \le t' \le t_{j+1} ) . $  
Note that, e.g., Barton \cite{barton} provides a useful compendium
of delta function representations. \\
iii) In problems where boundary effects are important,
$ G $ must, over each time interval, satisfy appropriate homogeneous
Dirichlet or Neumann (or possibly) mixed boundary conditions. 

Second, from (\ref{proc1adj}),
it is apparent that the adjoint equation associated with
the forward time evolution problem in (\ref{proc1})-(\ref{proc1b}) is of backward time form
in the forward time coordinate, $ t' $ 
(i.e., the signs on the diffusion and time derivative terms 
are of the same sign). Likewise, adjoint
equations associated with backward evolution equations
are of forward form in the backward time coordinate $ s' . $ Importantly,
while it is often difficult to determine analytical 
solutions to backward time problems \cite{pardoux,ma},
we can sometimes exploit the smallness of $ \Delta t' $
or $ \Delta s' $ to adapt relatively simple 
forward time solutions to the \textit{approximate} solution of 
associated backward problems. This point is illustrated in 
our solution of Burger's equation in Test Case 1.

Third, a variety of methods are available for computing non-incremental
Green's functions
\cite{barton, morsefeshbach}, all of which can be applied to
computing incremental solutions. 
Barton \cite{barton}
provides an accessible description of a number of approaches, including, e.g.,
eigenfunction expansions for problems
subject to boundary effects. 
We note in passing that numerical application of incremental Green's function solutions
has antecedents in the literature on the
boundary element method \cite{liu} and the so-called
Green element method \cite{tosaka, taigbenu}. The present framework (GFSM), 
however, 
differs from these in an essential way
due to its combined reliance on 
Green's function and stochastic process methods.

Fourth, when interpreting
incremental Green's functions as incremental
transition densities, 
incremental $ G $ should, strictly speaking, satisfy two 
consistency conditions: it must allow
recovery of the local mean drift, 
$ \mathbf{b} (\mathbf{x^{'}} , s^{'}) , $
and 
the local diffusion matrix, $ \mathbf{B} (\mathbf{x^{'}}, s^{'}) $
\begin{equation}\label{consist1}
\mathbf{b}(\mathbf{x^{'}}, s^{'}) =
\lim_{\Delta s^{'} \rightarrow 0} \frac{1}{\Delta s'}
\int_D ( \mathbf{x^{''}} - \mathbf{x^{'}}) 
G ( \mathbf{x^{'}}, s^{'}| \mathbf{x^{''}} , s^{'} + 
\Delta s^{'}) 
d \mathbf{x^{''}} 
\end{equation}
and
\begin{equation}\label{consist2}
B_{ij} (\mathbf{x^{'}}, s^{'}) = \lim_{\Delta s^{'} \rightarrow 0} 
\frac{1}{\Delta s'}
\int_D ( x_i^{''} - x_i^{'}) (x_j^{''} - x_j^{'}) 
G ( \mathbf{x^{'}}, s^{'}| \mathbf{x^{''}} , s^{'} + \Delta s^{'}) 
d \mathbf{x^{''}} , 
\end{equation}
where in this paper, $ B_{ij} = 2 \nu \delta_{ij} , $  
and where $ | \mathbf{x^{''}} - \mathbf{x^{'}}| < \epsilon , $
with $ \epsilon $ arbitrarily small
\cite{gardiner}.

%%%%%%%%%%%%%%%%%%%%%%%%%%%%%%%%%%%%%%%%%%%%%%%%%%%%%%%%%%

\end{document}